\documentclass[lettersize,journal]{IEEEtran}
\usepackage{amsmath,amsfonts}
\usepackage{algorithmic}
\usepackage{algorithm}
\usepackage[caption=false,font=normalsize,labelfont=sf,textfont=sf]{subfig}
\usepackage{textcomp}
\usepackage{stfloats}
\usepackage{verbatim}
\usepackage{graphicx}
\usepackage{cite}
\usepackage{tabularx}
\usepackage{rotating}
\usepackage{booktabs}
\usepackage{multirow}
\usepackage{afterpage}
\usepackage{svg}
\usepackage{pdflscape}
\usepackage{float}
\usepackage{xurl}      
\usepackage{enumitem}
\usepackage[letterpaper,top=2cm,bottom=2cm,left=3cm,right=3cm,marginparwidth=1.75cm]{geometry}
\usepackage{amsmath}
\usepackage{subfig}
\usepackage[colorlinks=true, allcolors=blue]{hyperref}
\usepackage{glossaries}
\newcommand{\etal}{\textit{et al.}}
\newglossaryentry{latex}{
  name=LaTeX,
  description={A typesetting system that is widely used for the production of scientific and mathematical documents due to its powerful handling of formulas and bibliographies}
} 
\newglossaryentry{}{
  name=,
  description={}
}

\newglossaryentry{VideoStreaming}{
  name={VideoStreaming},
  description={The process of transmitting video data over a network for real-time or on-demand viewing}
}

\newglossaryentry{Healthcare}{
  name={Healthcare},
  description={The maintenance or improvement of health through the prevention, diagnosis, treatment, and management of health impairments (e.g., wireless body area network)}
}

\newglossaryentry{SmartEnergy}{
  name={SmartEnergy},
  description={A sustainable and efficient energy management system (e.g., smart grid) that uses technology to monitor, control, and optimize energy consumption}
}

\newglossaryentry{IoT}{
  name={IoT},
  description={The Internet of Things refers to the network of interconnected devices, sensors, and objects that collect and exchange data to enable automation, monitoring, and control in various applications such as Cyber-Physical Systems (CPS)}
}

\newglossaryentry{IIoT}{
  name={IIoT},
  description={The Industrial Internet of Things (e.g., smart factory), which refers to the use of interconnected sensors, devices, and machines in industrial settings to improve efficiency, productivity, and safety}
}

\newglossaryentry{SmartCity}{
  name={SmartCity},
  description={A city that uses technology and data to enhance the quality of life for its residents, improve sustainability, and optimize resources and infrastructure (e.g., Bridge Structural Health Monitoring)}
}

\newglossaryentry{WeatherForecast}{
  name={WeatherForecast},
  description={The prediction of future weather conditions based on analysis of data and computer models}
}

\newglossaryentry{Drones}{
  name={Drones},
  description={Robots or unmanned aerial vehicles (UAVs) used for various purposes including aerial photography, surveillance, delivery, and mapping}
}

\newglossaryentry{CrowdSourcing}{
  name={Crowdsourcing},
  description={The practice of obtaining services or content by soliciting contributions from a large group of devices, typically via the internet (e.g., crowd sensing)}
}

\newglossaryentry{MobileComputing}{
  name={MobileComputing},
  description={Mobile Computing (MCO) refers to the use of mobile devices and wireless networks to access and share information and resources from anywhere at any time}
}

\newglossaryentry{5G}{
  name={5G},
  description={The fifth generation of wireless technology for cellular networks, designed to provide faster data speeds, lower latency, and increased connectivity for a wide range of devices and applications}
}

\newglossaryentry{6G}{
  name={6G},
  description={The sixth generation of wireless technology, still in development, expected to offer faster data speeds, lower latency, and new capabilities such as holographic communication and advanced AI}
}

\newglossaryentry{SmartTransport}{
  name={Smart Transport},
  description={The integration of technology and data analytics into transportation systems to improve efficiency, safety, and sustainability, and provide better mobility options for users (i.e., Intelligent Transportation System, Vehicle to Grid, maritime communication networks)}
}

\newglossaryentry{SupplyChain}{
  name={SupplyChain},
  description={The network of interconnected entities involved in the production, distribution, and delivery of goods and services from suppliers to consumers}
}

\newglossaryentry{ECommerce}{
  name={ECommerce},
  description={Electronic commerce, the buying and selling of goods and services over the internet, encompassing online retail, digital payments, and other related activities}
}

\newglossaryentry{SmartFarm}{
  name={SmartFarm},
  description={The use of technology, data analytics, and automation in agricultural practices to improve productivity, sustainability, and resource efficiency}
}

\newglossaryentry{SmartWater}{
  name={SmartWater},
  description={The use of technology and data analytics to monitor, manage, and optimize water resources, infrastructure, and distribution systems for efficiency and sustainability}
}

\newglossaryentry{Surveillance}{
  name={Surveillance},
  description={The monitoring and observation of activities, behavior, or other data for the purpose of gathering information, ensuring security, or enforcing regulations (e.g., public safety, Bridge Structural Health Monitoring)}.
}

\newglossaryentry{SmartCameras}{
  name={SmartCameras},
  description={A surveillance camera equipped with advanced features such as motion detection, facial recognition, and remote monitoring capabilities, often used for security and monitoring purposes}
}

\newglossaryentry{SmartHome}{
  name={SmartHome},
  description={A residence equipped with interconnected devices and systems that automate and control various functions such as lighting, heating, security, and entertainment, for improved comfort, convenience, and energy efficiency}
}

\newglossaryentry{DataSharing}{
  name={Data Sharing},
  description={The practice of sharing data between individuals, organizations, or systems for various purposes such as collaboration, analysis, or decision-making}
}

\newglossaryentry{DataTrading}{
  name={Data Trading},
  description={The buying and selling of data between parties, often for commercial purposes such as marketing, research, or analytics}
}

\newglossaryentry{Performance}{
  name={Performance},
  description={The improvement of how well a system, process, or component accomplishes its intended function or goal, often assessed in terms of speed, efficiency, or effectiveness (e.g., improvement of system throughput, availability, storage efficiency, communication overhead, etc)}
}

\newglossaryentry{Trust}{
  name={Trust},
  description={The belief or confidence in the reliability, integrity, and competence of a person, organization, or system, often based on past experiences or reputation}
}

\newglossaryentry{Privacy}{
  name={Privacy},
  description={The right to control access to personal information and the ability to keep sensitive data confidential, often regulated by laws and policies to protect individuals' rights}
}

\newglossaryentry{Security}{
  name={Security},
  description={The protection of data, systems, and resources from unauthorized access, disclosure, alteration, destruction, or disruption, often achieved through various security measures and protocols such as access control and confidentiality}
}

\newglossaryentry{Transparency}{
  name={Transparency},
  description={The openness and clarity in decision-making, actions, or operations, allowing stakeholders to understand and assess the processes, motivations, and outcomes involved}
}

\newglossaryentry{problemReliability}{
  name={Reliability},
  description={The consistency and dependability of a system or component to perform its intended function accurately and predictably under various conditions, often measured in terms of uptime, failure rate, or mean time between failures}
}

\newglossaryentry{problemScalability}{
  name={Scalability},
  description={The ability of a system, process, or architecture to handle increasing workload or data volume without significant performance degradation or the need for major changes or reconfiguration}
}

\newglossaryentry{Incentivization}{
  name={Incentivization},
  description={The use of incentives or rewards to motivate individuals or devices to perform certain actions or behaviors, often used in contexts such as economics, marketing, or behavior change}
}

\newglossaryentry{Programmability}{
  name={Programmability},
  description={The ability of a system, device, or platform to be programmed or customized to perform specific tasks or functions, through software development or configuration}
}

\newglossaryentry{Experimentation}{
  name={Experimentation},
  description={The development of tools, platforms, or simulation environments aimed at facilitating research activities tailored to the research domain such as resource monitoring tools}
}

\newglossaryentry{DataProvenance}{
  name={DataProvenance},
  description={The history and origin of data, including its creation, transformation, ownership, and movement, often tracked and recorded to ensure data quality, integrity, and compliance}
}

\newglossaryentry{DataRecovery}{
  name={DataRecovery},
  description={The process of restoring lost, damaged, or corrupted data from backup or secondary storage systems, often performed in response to data loss incidents such as hardware failures, human errors, or cyberattacks}
}

\newglossaryentry{Model}{
  name={Model},
  description={A representation or abstraction designed to simulate, describe, or analyze a system, process, or phenomenon such as Mathematical, Computational, and Statistical Models}
}
\newglossaryentry{Framework}{
  name={Framework},
  description={An organized structure or set of tools that provides a foundation for developing software applications or addressing specific challenges}
}
\newglossaryentry{Platform}{
  name={Platform},
  description={A computing environment that serves as a base for developing, executing, and managing software applications}
}
\newglossaryentry{Architecture}{
  name={Architecture},
  description={The overall design and structure of a system or software, including its components, modules, and their relationships (e.g., block format in a Blockchain platform)}
}
\newglossaryentry{Methodology}{
  name={Methodology},
  description={A systematic approach or set of principles and processes used to conduct research, analysis, or problem-solving in a particular field}
}
\newglossaryentry{Algorithm}{
  name={Algorithm},
  description={A step-by-step procedure or set of rules for solving a specific problem or performing a particular computation}
}
\newglossaryentry{Protocol}{
  name={Protocol},
  description={A set of rules defining the format and sequence of messages exchanged between devices or entities in a communication network (e.g., consensus protocol}
}
\newglossaryentry{SoftwareArtifact}{
  name={SoftwareArtifact},
  description={A tangible, software-related object or item resulting from the software development process, such as code, or executable files (e.g., smart contract, middleware)}
}
\newglossaryentry{Service}{
  name={Service},
  description={A function or set of functions provided by a system, software, or platform, often accessible over a network}
}
\newglossaryentry{Mechanism}{
  name={Mechanism},
  description={A device or process designed to perform a specific function, often within a larger system or framework}
}

\newglossaryentry{Reinforcement}{
  name={Reinforcement},
  description={Reinforcement learning is a machine learning approach where an agent learns to make decisions by interacting with an environment and receiving feedback as rewards or penalties (e.g., deep reinforcement learning)}
}

\newglossaryentry{Federated}{
  name={Federated},
  description={Federated learning where model training is performed on decentralized devices, sharing only model updates with a central server}
}

\newglossaryentry{Unsupervised}{
  name={Unsupervised},
  description={Unsupervised machine learning where a model learns patterns in data without labeled examples}
}

\newglossaryentry{Supervised}{
  name={Supervised},
  description={Supervised machine learning where a model learns from labeled examples to make predictions or classify data}
}

\newglossaryentry{Heuristic}{
  name={Heuristic},
  description={A problem-solving approach using practical rules or strategies to find approximate solutions (e.g., Greedy algorithms)}
}

\newglossaryentry{Metaheuristic}{
  name={Metaheuristic},
  description={A problem-solving strategy guiding the search for optimal solutions across diverse domains (e.g., Genetic Algorithm, Particle Swarm Optimization)}
}

\newglossaryentry{aiMatching}{
  name={Matching},
  description={The process of pairing or aligning data points, entities, or features based on specific criteria to achieve a particular objective}
}

\newglossaryentry{Reasoning}{
  name={Reasoning},
  description={Drawing logical inferences or reaching conclusions based on available information (e.g., Fuzzy logical inference}
}

\newglossaryentry{Authentication}{
  name={Authentication},
  description={The process of verifying the identity of a user or entity, typically through the presentation of credentials such as passwords or cryptographic keys}
}

\newglossaryentry{Encryption}{
  name={Encryption},
  description={The process of encoding data in such a way that only authorized parties can access and understand it, typically using cryptographic algorithms and keys}
}

\newglossaryentry{AccessControl}{
  name={AccessControl},
  description={The process of restricting or granting access to resources or services based on policies or rules defining who is allowed to access what, when, and under what conditions}
}

\newglossaryentry{Verifiability}{
  name={Verifiability},
  description={The ability to verify the correctness, integrity, or authenticity of data, transactions, or processes using cryptographic techniques or other methods}
}

\newglossaryentry{TrustManagement}{
  name={TrustManagement},
  description={Trust Management is the process of establishing, maintaining, and evaluating trust relationships between entities in a system, often involving reputation mechanisms}
}

\newglossaryentry{Credibility}{
  name={Credibility},
  description={The believability or reliability of information, systems, or entities, often based on factors such as reputation, authority, and evidence}
}

\newglossaryentry{Integrity}{
  name={Integrity},
  description={The assurance that data or systems have not been tampered with or altered in an unauthorized manner, ensuring accuracy, consistency, and trustworthiness}
}

\newglossaryentry{SecurityAvailability}{
  name={Availability},
  description={The assurance that data, systems, or services are accessible and usable when needed, typically achieved through redundancy, fault tolerance, and recovery measures}
}

\newglossaryentry{Immutability}{
  name={Immutability},
  description={The property of data or records that prevents them from being modified or deleted once they are created, ensuring data integrity and auditability}
}

\newglossaryentry{Confidentiality}{
  name={Confidentiality},
  description={The protection of sensitive or private information from unauthorized access, disclosure, or exposure}
}

\newglossaryentry{Authorization}{
  name={Authorization},
  description={The process of granting or denying permissions to access specific resources or perform certain actions, based on the identity and privileges of the requester}
}

\newglossaryentry{NonRepudiation}{
  name={NonRepudiation},
  description={The ability to prove the origin or authenticity of a communication or transaction and prevent the sender from denying their involvement or actions, often achieved through digital signatures and audit trails}
}

\newglossaryentry{Anonymity}{
  name={Anonymity},
  description={The state of being anonymous or unidentifiable, where the identity or personal information of individuals is not known or disclosed}
}

\newglossaryentry{DiffPrivacy}{
  name={DiffPrivacy},
  description={Differential Privacy is a privacy-preserving data analysis technique that ensures statistical queries do not reveal information about individual data points}
}

\newglossaryentry{OPRF}{
  name={OPRF},
  description={Oblivious Pseudorandom Function (OPRF) is a cryptographic primitive that allows a system to compute pseudorandom values for an entity without learning anything about the entity's input}
}

\newglossaryentry{PSI}{
  name={PSI},
  description={Private Set Intersection (PSI) is a cryptographic protocol that allows two parties to compute the intersection of their sets without revealing any information about elements that are not in the intersection}
}

\newglossaryentry{HomoEncryption}{
  name={HomoEncryption},
  description={Homomorphic Encryption is a form of encryption that allows mathematical operations to be performed on encrypted data without decrypting it first, preserving privacy while enabling computation on sensitive data}
}

\newglossaryentry{RingSignature}{
  name={RingSignature},
  description={Ring Signature is a digital signature scheme that allows a user to sign a message on behalf of a group without revealing which member of the group performed the signature (e.g., Threshold Ring Signature)}
}

\newglossaryentry{ZeroKnowlProof}{
  name={ZeroKnowlProof},
  description={Zero-Knowledge Proof is a cryptographic protocol that allows one party (the prover) to prove to another party (the verifier) that they possess certain knowledge or information without revealing the actual knowledge or information}
}

\newglossaryentry{Pseudonymity}{
  name={Pseudonymity},
  description={The use of pseudonyms or aliases to conceal the true identity of individuals while still allowing them to be uniquely identified within a system or context}
}

\newglossaryentry{BlindSignature}{
  name={BlindSignature},
  description={Blind Signature is a digital signature scheme that allows a user to obtain a valid signature on a message without revealing the content of the message to the signer}
}

\newglossaryentry{SecretSharing}{
  name={SecretSharing},
  description={Secret Sharing is a cryptographic technique that divides a secret into multiple shares distributed among participants, requiring a threshold number of shares to reconstruct the original secret}
}

\newglossaryentry{AutoEncoder}{
  name={AutoEncoder},
  description={A type of neural network architecture used for unsupervised learning that learns to encode input data into a compact representation and decode it back to its original form}
}

\newglossaryentry{HybridIdentity}{
  name={HybridIdentity},
  description={A form of digital identity that combines attributes from multiple sources or domains, often used to provide more robust and flexible authentication and authorization mechanisms}
}

\newglossaryentry{Trading}{
  name={Trading},
  description={The exchange of data, resources, or services between parties in return for something of value, such as money, goods, or other assets}
}

\newglossaryentry{Offloading}{
  name={Offloading},
  description={The process of transferring data, tasks, or computational load from one system or device to another, often to optimize performance, efficiency, or resource utilization}
}

\newglossaryentry{Balancing}{
  name={Balancing},
  description={The distribution of computational or network workload across multiple resources or servers to ensure optimal resource utilization, performance, and reliability}
}

\newglossaryentry{Placement}{
  name={Placement},
  description={The Placement is the strategic positioning or deployment of services or applications within a network or infrastructure to optimize performance, availability, and scalability}
}

\newglossaryentry{scaling}{
  name={Scaling},
  description={The automatic adjustment of resources or capacity in response to changing workload demands, typically to maintain optimal performance and cost efficiency}
}

\newglossaryentry{Scheduling}{
  name={Scheduling},
  description={The process of allocating resources, tasks, or activities over time, often to optimize efficiency, minimize delays, or meet deadlines}
}

\newglossaryentry{allocMatching}{
  name={Matching},
  description={The matching approach in resource allocation is a method used to efficiently distribute resources by matching demand with available supply based on certain criteria or preferences}
}

\newglossaryentry{CompDelay}{
  name={CompDelay},
  description={Computation Delay (i.e., service delay) refers to the amount of time it takes for a service to be processed which includes queuing delay and execution time}
}

\newglossaryentry{Throughput}{
  name={Throughput},
  description={The rate at which data or tasks are processed, transmitted, or completed within a system or network, including data processing speed and cache hit rate}
}

\newglossaryentry{SecurityVulner}{
  name={SecurityVulner},
  description={Security Vulnerability is a metric that measures a weakness or flaw in a system or application that could be exploited by attackers to compromise the confidentiality, integrity, or availability of data or resources}
}

\newglossaryentry{Latency}{
  name={Latency},
  description={The time delay between the initiation of a request or action and the response or result, often caused by factors such as processing time, network congestion, or communication delay}
}

\newglossaryentry{GasConsumption}{
  name={GasConsumption},
  description={In the context of blockchain technology, gas consumption refers to the amount of computational resources consumed by executing smart contracts or transactions on a blockchain network, measured in units of "gas."}
}

\newglossaryentry{UtilSocialWelfare}{
  name={UtilSocialWelfare},
  description={Social Welfare Utility is a measure of the overall well-being or utility of society, often influenced by factors such as income distribution, access to resources, and social policies}
}

\newglossaryentry{ConvergeSpeed}{
  name={ConvergeSpeed},
  description={Convergence Speed measures the rate at which a system, algorithm, or process converges to a stable or optimal solution, often measured in iterations, epochs, or time units}
}

\newglossaryentry{CommCost}{
  name={CommCost},
  description={Communication Cost is the resources required to transmit data or messages between entities in a system or network, often influenced by factors such as bandwidth and latency}
}

\newglossaryentry{CompCost}{
  name={CompCost},
  description={Computation Cost is the resources required to perform computations or processing tasks in a system or algorithm, often influenced by factors such as algorithm complexity, hardware specifications, and workload size}
}

\newglossaryentry{Accuracy}{
  name={Accuracy},
  description={The degree of correctness or precision of a measurement, calculation, or prediction (specifically in AI methods), often expressed as the ratio of correct outcomes to total outcomes}
}

\newglossaryentry{PowerConsumption}{
  name={PowerConsumption},
  description={The amount of electrical power consumed by a system, device, or component}
}

\newglossaryentry{EnergyConsumption}{
  name={EnergyConsumption},
  description={The amount of energy consumed by a system, device, or process over a specific period}
}

\newglossaryentry{CommUtil}{
  name={CommUtil},
  description={Communication Utilization is the degree to which communication resources, such as network bandwidth, channels, or protocols, are utilized or allocated}
}

\newglossaryentry{CompUtil}{
  name={CompUtil},
  description={Computation Utilization measures the degree to which computational resources, such as CPU, memory, or processing units, are utilized or allocated}
}

\newglossaryentry{metricScalability}{
  name={Scalability},
  description={The ability of a system, architecture, or technology to handle increasing workload demands or user interactions without sacrificing performance, stability, or quality of service}
}

\newglossaryentry{StorageUtil}{
  name={StorageUtil},
  description={Storage Utilization refers to the measure of how effectively storage resources are being used within a system or infrastructure}
}

\newglossaryentry{MetricReliability}{
  name={Reliability},
  description={The ability of a system, component, or process to consistently perform its intended function or deliver expected results under specified conditions, often measured in terms of uptime, failure rate, or mean time between failures (MTBF)}
}

\newglossaryentry{MetricAvailability}{
  name={Availability},
  description={The percentage of time that a system, service, or resource is operational and accessible for use, often expressed as a fraction or percentage of total uptime over a given period}
}

\newglossaryentry{Precision}{
  name={Precision},
  description={The measure of the proportion of true positive results among all positive results returned by an algorithm}
}

\newglossaryentry{Recall}{
  name={Recall},
  description={The measure of the proportion of true positive results among all actual positive instances}
}

\newglossaryentry{F1Score}{
  name={F1Score},
  description={The harmonic mean of precision and recall, providing a single metric that balances both precision and recall in classification or detection tasks}
}

\newglossaryentry{StorageCost}{
  name={StorageCost},
  description={The expenses or resources required to store and maintain data or information in a system or storage infrastructure, often influenced by factors such as capacity, redundancy, and access frequency}
}

\newglossaryentry{FailureCost}{
  name={FailureCost},
  description={The direct and indirect costs associated with system failures, downtime, or errors, including lost revenue, productivity, reputation damage, and recovery expenses}
}

\newglossaryentry{metricPrivacy}{
  name={Privacy},
  description={The degree to which sensitive or personal information is protected from unauthorized access, disclosure, or misuse}
}

\newglossaryentry{LinesOfCode}{
  name={LinesOfCode},
  description={Lines of Code is a measure of the size or complexity of a software project, often used to estimate development effort, maintenance costs, and code quality}
}

\newglossaryentry{Reputation}{
  name={Reputation},
  description={The collective perception, opinion, or evaluation of an entity's character, or trustworthiness, often based on past behavior, interactions, or feedback from others}
}

\newglossaryentry{WinningRate}{
  name={WinningRate},
  description={The percentage or proportion of successful outcomes or victories achieved in a given context, such as block mining, game, or trading}
}

\newglossaryentry{Jitter}{
  name={Jitter},
  description={The variation or deviation in the latency or timing of data packets or events in a network, often caused by congestion, queuing delays, or fluctuating network conditions}
}

\newglossaryentry{IPFS}{
  name={IPFS},
  description={InterPlanetary File System (IPFS) is a protocol and network designed to create a peer-to-peer method of storing in and sharing a distributed file system}
}

\newglossaryentry{RFID}{
  name={RFID},
  description={Radio-Frequency Identification (RFID) is a technology that uses electromagnetic fields to automatically identify and track tags attached to objects}
}

\newglossaryentry{CellularNet}{
  name={CellularNet},
  description={Cellular Networks refer to wireless communication networks that provide voice and data services to mobile users}
}

\newglossaryentry{MEC}{
  name={MEC},
  description={Multi-access Edge Computing (MEC) is a network architecture that enables cloud computing capabilities and an IT service environment at the edge of the cellular network}
}

\newglossaryentry{Docker}{
  name={Docker},
  description={Docker is a platform for developing, shipping, and running applications inside containers}
}

\newglossaryentry{InfluxDB}{
  name={InfluxDB},
  description={InfluxDB is a time-series database designed to handle high write and query loads for monitoring, analytics, and IoT applications}
}

\newglossaryentry{Tech5G}{
  name={5G},
  description={5G is the fifth generation of cellular network technology, designed to provide faster data speeds, lower latency, and greater capacity than previous generations}
}

\newglossaryentry{Tech6G}{
  name={6G},
  description={6G is the sixth generation of cellular network technology, envisioned to further enhance the capabilities of wireless communication systems beyond 5G}
}

\newglossaryentry{WiFi}{
  name={WiFi},
  description={WiFi is a technology that allows electronic devices to connect to a wireless local area network (WLAN)}
}

\newglossaryentry{Bluetooth}{
  name={Bluetooth},
  description={Bluetooth is a wireless technology standard for exchanging data over short distances using short-wavelength radio waves}
}

\newglossaryentry{Zigbee}{
  name={Zigbee},
  description={Zigbee is a low-power, low-data-rate wireless communication protocol used for small-scale projects that require wireless connection}
}

\newglossaryentry{ANT}{
  name={ANT},
  description={ANT is a proprietary wireless sensor network technology developed by Garmin Ltd. ANT}
}

\newglossaryentry{NFC}{
  name={NFC},
  description={Near Field Communication (NFC) is a short-range wireless technology that allows communication between electronic devices by bringing them close together}
}

\newglossaryentry{SDN}{
  name={SDN},
  description={Software-Defined Networking (SDN) is an architecture approach that decouples the control plane from the data plane, allowing centralized control and programmability of network devices}
}

\newglossaryentry{NFV}{
  name={NFV},
  description={Network Functions Virtualization (NFV) is a network architecture concept that uses virtualization technologies to virtualize network functions, such as firewalls, routers, and load balancers}
}

\newglossaryentry{LightNet}{
  name={LightNet},
  description={The Lightning Network is a decentralized scaling solution for blockchain networks, designed to enable high-frequency transactions off-chain, thereby reducing congestion and increasing scalability}
}

\newglossaryentry{ETHTestnet}{
  name={ETHTestnet},
  description={The Ethereum Testnet encompasses various test networks designed for developers to experiment with Ethereum blockchain applications, including Ropsten, Rinkeby, Reth, and TestRPC}
}

\newglossaryentry{Redis}{
  name={Redis},
  description={Redis is an open-source, in-memory data store used as a database, cache, and message broker}
}

\newglossaryentry{OpenStack}{
  name={OpenStack},
  description={OpenStack is a cloud computing platform that provides infrastructure-as-a-service (IaaS) for deploying and managing virtualized resources}
}

\newglossaryentry{CouchDB}{
  name={CouchDB},
  description={CouchDB is a distributed document-oriented NoSQL database that uses JSON to store data}
}

\newglossaryentry{Kafka}{
  name={Kafka},
  description={Apache Kafka is an open-source stream-processing software platform developed by the Apache Software Foundation}
}

\newglossaryentry{BigchainDB}{
  name={BigchainDB},
  description={BigchainDB is a blockchain database designed for storing and querying large volumes of structured data}
}

\newglossaryentry{GoETH}{
  name={GoETH},
  description={Go Ethereum (Geth) is one of the three original implementations of the Ethereum protocol, used to interact with the Ethereum blockchain}
}

\newglossaryentry{RaspberryPi}{
  name={RaspberryPi},
  description={Raspberry Pi is a series of small single-board computers commonly used for educational and embedded computing projects, including Nano Pi, LattePanda, and Orange Pi}
}

\newglossaryentry{Nodejs}{
  name={Nodejs},
  description={Node.js is an open-source, cross-platform JavaScript runtime environment commonly used for server-side programming}
}

\newglossaryentry{Mininet}{
  name={Mininet},
  description={Mininet is a software emulator for creating virtual networks on a single machine, enabling the testing and development of software-defined networking (SDN) applications}
}

\newglossaryentry{REST}{
  name={REST},
  description={Representational State Transfer (REST) is a software architectural style that defines a set of constraints for creating web services}
}

\newglossaryentry{Arduino}{
  name={Arduino},
  description={Arduino is an open-source hardware and software platform used for building electronic projects and prototypes, consisting of microcontroller boards and a development environment}
}

\newglossaryentry{NodeRED}{
  name={NodeRED},
  description={Node-RED is a flow-based programming tool for wiring together hardware devices, APIs, and online services, enabling the rapid development of IoT applications}
}

\newglossaryentry{iFogSim}{
  name={iFogSim},
  description={iFogSim is a simulation toolkit for modeling and simulating fog computing environments and applications}
}

\newglossaryentry{Brain4Net}{
  name={Brain4Net},
  description={Brain4Net is a research project focused on developing innovative solutions for next-generation communication networks and internet infrastructures}
}

\newglossaryentry{EdisonSoC}{
  name={EdisonSoC},
  description={Intel Edison SoC is a small, low-power system-on-module (SoM) developed by Intel, designed for IoT and embedded applications}
}

\newglossaryentry{Wireshark}{
  name={Wireshark},
  description={Wireshark is a free and open-source packet analyzer used for network troubleshooting, analysis, and protocol development}
}

\newglossaryentry{Jolinar}{
  name={Jolinar},
  description={Jolinar is an energy measurement tool designed to provide accurate, user-friendly insights into software energy consumption}
}

\newglossaryentry{GPS}{
  name={GPS},
  description={Global Positioning System (GPS) is a satellite-based navigation system that provides location and time information to users}
}

\newglossaryentry{Contiki}{
  name={Contiki},
  description={Contiki is an open-source operating system designed for the Internet of Things (IoT) devices and embedded systems}
}

\newglossaryentry{OPNET}{
  name={OPNET},
  description={OPNET (Optimized Network Engineering Tool) is a network simulation software suite used for modeling, analyzing, and optimizing communication networks and systems}
}

\newglossaryentry{GMapAPI}{
  name={GMapAPI},
  description={Google Maps API (GMapAPI) is a set of application programming interfaces (APIs) provided by Google for accessing and integrating Google Maps functionality into web and mobile applications}
}

\newglossaryentry{Chainlink}{
  name={Chainlink},
  description={Chainlink is a decentralized oracle network that enables smart contracts on various blockchain platforms to securely interact with external data sources, APIs, and payment systems}
}

\newglossaryentry{NS3}{
  name={NS3},
  description={ns-3 (Network Simulator version 3) is a discrete-event network simulator used for simulating and analyzing communication networks and protocols}
}

\newglossaryentry{Postman}{
  name={Postman},
  description={A popular API testing tool used for sending HTTP requests and examining responses during the development and testing of APIs}
}

\newglossaryentry{BlackPill}{
  name={BlackPill},
  description={Black Pill is a compact development board featuring the STM32 microcontroller, known for its small size and enhanced computational capabilities compared to Arduino boards}
}

\newglossaryentry{HealthShield}{
  name={HealthShield},
  description={E-Health Shield is a biometric sensor platform for Arduino and Raspberry Pi, facilitating medical parameter monitoring for patients.}
}

\newglossaryentry{JMeter}{
  name={JMeter},
  description={An open-source tool used for performance testing of web applications}
}

\newglossaryentry{OpenThread}{
  name={OpenThread},
  description={Open Thread is an open-source implementation of the Thread networking protocol, which is a networking technology designed for smart home and Internet of Things (IoT) applications}
}

\newglossaryentry{AVISPA}{
  name={AVISPA},
  description={AVISPA is a push-button tool for the automated validation of Internet security-sensitive protocols and applications}
}

\newglossaryentry{Onos}{
  name={Onos},
  description={Open Network Operating System (ONOS), a software-defined networking operating system designed to manage network devices and traffic}
}

\newglossaryentry{Prometheus}{
  name={Prometheus},
  description={An open-source monitoring and alerting toolkit often used for monitoring containerized applications in cloud-native environments}
}

\newglossaryentry{NOMA}{
  name={NOMA},
  description={Non-Orthogonal Multiple Access (NOMA), a technology used in wireless communication systems to improve spectral efficiency and capacity}
}

\newglossaryentry{RoF}{
  name={RoF},
  description={Radio over Fiber, a technology that involves the transmission of radio signals over optical fibers for long-distance communication}
}

\newglossaryentry{SQLite}{
  name={SQLite},
  description={A lightweight, embedded relational database management system (RDBMS) commonly used in mobile apps and small-scale applications}
}

\newglossaryentry{FreeStyleLib}{
  name={FreeStyleLib},
  description={Abbott FreeStyle Libre is a continuous glucose monitoring system that utilizes advanced sensors for people managing diabetes}
}

\newglossaryentry{OrbitDB}{
  name={OrbitDB},
  description={A decentralized database built on top of the IPFS (InterPlanetary File System), designed for peer-to-peer applications}
}

\newglossaryentry{Ganache}{
  name={Ganache},
  description={A personal blockchain for Ethereum development, providing a local environment for testing and development purposes}
}

\newglossaryentry{OneSwarm}{
  name={OneSwarm},
  description={A privacy-preserving peer-to-peer file-sharing tool designed to protect users' anonymity while sharing files}
}

\newglossaryentry{RemixIDE}{
  name={RemixIDE},
  description={An integrated development environment (IDE) for Solidity, the programming language used for writing smart contracts on the Ethereum blockchain}
}

\newglossaryentry{JetXavNx}{
  name={JetXavNx},
  description={The Jetson Xavier NX is a compact and powerful AI computing module designed by NVIDIA for edge computing applications, offering high performance and energy efficiency}
}

\newglossaryentry{CIDDS}{
  name={CIDDS},
  description={Cyber-attack Intrusion Detection Dataset, a dataset commonly used for evaluating intrusion detection systems' effectiveness}
}

\newglossaryentry{PostgreSQL}{
  name={PostgreSQL},
  description={An open-source relational database management system known for its reliability, robustness, and advanced features}
}

\newglossaryentry{OMNeT}{
  name={OMNeT},
  description={OMNeT++, a modular, component-based simulation environment used for building network simulators}
}

\newglossaryentry{SQL}{
  name={SQL},
  description={Microsoft SQL Server is a relational database management system (RDBMS) developed by Microsoft}
}

\newglossaryentry{MYSQL}{
  name={MYSQL},
  description={An open-source relational database management system known for its speed, reliability, and ease of use}
}

\newglossaryentry{SUMO}{
  name={SUMO},
  description={Simulation of Urban MObility (SUMO), an open-source traffic simulation suite used for modeling and simulating urban traffic scenarios}
}

\newglossaryentry{K8S}{
  name={K8S},
  description={Kubernetes, an open-source container orchestration platform used for automating the deployment, scaling, and management of containerized applications}
}

\newglossaryentry{LevelDB}{
  name={LevelDB},
  description={A fast and lightweight key-value storage library developed by Google, optimized for performance and efficiency}
}

\newglossaryentry{JMTSim}{
  name={JMTSim},
  description={JMT (Java Modeling Tools) is an open-source suite of tools such as JSIM (Java Simulator) for performance modeling and simulation of computer and communication systems}
}

\newglossaryentry{FoBSim}{
  name={FoBSim},
  description={An open-source simulation tool for integrated fog-blockchain systems}
}

\newglossaryentry{IntelSGX}{
  name={IntelSGX},
  description={Intel Software Guard Extensions, a set of security-related instruction codes built into Intel CPUs to enhance application security through hardware-based encryption}
}

\newglossaryentry{JetsonNano}{
  name={JetsonNano},
  description={The NVIDIA Jetson Nano Developer Kit is a platform designed for developing AI applications and deploying edge computing solutions}
}

\newglossaryentry{ZooKeeper}{
  name={ZooKeeper},
  description={A centralized service for maintaining configuration information, naming, providing distributed synchronization, and providing group services}
}

\newglossaryentry{Caliper}{
  name={Caliper},
  description={Caliper is a blockchain benchmarking tool developed within the Hyperledger project}
}

\newglossaryentry{NoPoverty}{
  name={NoPoverty},
  description={The goal to end poverty in all its forms everywhere, ensuring that all people have equal rights to economic resources, access to basic services, ownership, and control over land and other forms of property}
}

\newglossaryentry{ZeroHunger}{
  name={ZeroHunger},
  description={The goal to end hunger, achieve food security and improved nutrition, and promote sustainable agriculture, ensuring that everyone has access to safe, nutritious, and sufficient food all year round}
}

\newglossaryentry{HealthWellbeing}{
  name={HealthWellbeing},
  description={The goal of Good Health and Well-being is to ensure healthy lives and promote well-being for all at all ages}
}

\newglossaryentry{EducationQuality}{
  name={EducationQuality},
  description={The goal to ensure inclusive and equitable quality education and promote lifelong learning opportunities for all}
}

\newglossaryentry{GenderEquality}{
  name={GenderEquality},
  description={The goal to achieve gender equality and empower all women and girls, and ensure equal access to education, healthcare, and economic opportunities}
}

\newglossaryentry{WaterSanitation}{
  name={WaterSanitation},
  description={The goal of Clean Water and Sanitation is to ensure availability and sustainable management of water and sanitation for all}
}

\newglossaryentry{Energy}{
  name={Energy},
  description={The goal of Affordable and Clean Energy is to ensure access to affordable, reliable, sustainable, and modern energy for all, aiming to promote renewable energy sources, increase energy efficiency, and expand energy infrastructure and technology in developing countries}
}

\newglossaryentry{WorkEconomic}{
  name={WorkEconomic},
  description={Decent Work and Economic Growth aims to promote sustained, inclusive, and sustainable economic growth, full and productive employment, and decent work for all}
}

\newglossaryentry{IndustryStructure}{
  name={IndustryStructure},
  description={Industry, Innovation, and Infrastructure goal aims to build resilient infrastructure, promote inclusive and sustainable industrialization, and foster innovation, aiming to enhance technological progress, upgrade infrastructure and industries, and support small-scale enterprises and innovation hubs}
}

\newglossaryentry{ReduceInequality}{
  name={ReducedInequality},
  description={The goal to reduce inequality within and among countries, aiming to empower and promote the social, economic, and political inclusion of all}
}

\newglossaryentry{SustainableCity}{
  name={SustainableCity},
  description={Sustainable Cities and Communities aims to make cities and human settlements inclusive, safe, resilient, and sustainable, aiming to ensure access to safe and affordable housing, basic services, and transportation, and reduce the environmental impact of urbanization}
}

\newglossaryentry{ResponsibleConsProd}{
  name={ResponsibleConsProd},
  description={Responsible Consumption and Production aims to ensure sustainable consumption and production patterns, aiming to achieve sustainable management and efficient use of natural resources, reduce waste generation, and minimize environmental degradation throughout the lifecycle of products and services}
}

\newglossaryentry{ClimateAction}{
  name={ClimateAction},
  description={The goal to take urgent action to combat climate change and its impacts, aiming to strengthen resilience and adaptive capacity to climate-related hazards and natural disasters, and integrate climate change measures into national policies, strategies, and planning}
}

\newglossaryentry{LifeBelowWater}{
  name={LifeBelowWater},
  description={The goal to conserve and sustainably use the oceans, seas, and marine resources for sustainable development}
}

\newglossaryentry{LifeOnLand}{
  name={LifeOnLand},
  description={The goal to protect, restore, and promote sustainable use of terrestrial ecosystems, halt and reverse land degradation, and biodiversity loss}
}

\newglossaryentry{PeaceJustice}{
  name={PeaceJustice},
  description={The goal Peace, Justice, and Strong Institutions aims to promote peaceful and inclusive societies for sustainable development, provide access to justice for all, and build effective, accountable, and inclusive institutions at all levels}
}

\newglossaryentry{GlobalPartnership}{
  name={GlobalPartnership},
  description={The goal to strengthen the means of implementation and revitalize the global partnership for sustainable development}
}

\newglossaryentry{MQTT}{
  name={MQTT},
  description={MQTT (Message Queuing Telemetry Transport) is a lightweight messaging protocol commonly employed in Internet of Things (IoT) applications for efficient machine-to-machine (M2M) communication in constrained environments}
}

\newglossaryentry{VANET}{
  name={VANET},
  description={VANET (Vehicular Ad Hoc Network) is a type of ad hoc network used to enable communication between vehicles (V2V) and between vehicles and roadside infrastructure (V2I)}
}

\newglossaryentry{TLS}{
  name={TLS},
  description={TLS (Transport Layer Security) is a cryptographic protocol used to secure communication over a computer network}
}

\newglossaryentry{M2M}{
  name={M2M},
  description={M2M (Machine-to-Machine) refers to direct communication between devices or machines}
}

\newglossaryentry{CAN}{
  name={CAN},
  description={CAN (Controller Area Network) is a robust serial communication protocol commonly used in automotive and industrial applications}
}

\newglossaryentry{OFDM}{
  name={OFDM},
  description={OFDM (Orthogonal Frequency Division Multiplexing) is a digital modulation technique used in wireless communication systems to transmit data over multiple subcarriers simultaneously}
}

\newglossaryentry{SoAP}{
  name={SoAP},
  description={SoAP (Simple Object Access Protocol) is a protocol for exchanging structured information in the implementation of web services}
}

\newglossaryentry{CoAP}{
  name={CoAP},
  description={CoAP (Constrained Application Protocol) is a lightweight protocol designed for constrained devices and networks in the context of the Internet of Things (IoT)}
}

\newglossaryentry{STEP}{
  name={STEP},
  description={Service and Topology Exchange Protocol (STEP) enables service availability and network scalability through centralized repository management, enhancing network administration efficiency}
}

\newglossaryentry{IE1609.2}{
  name={IE1609.2},
  description={IEEE 1609.2 is a standard defining security services and protocols for wireless access in vehicular environments (WAVE)}
}

\newglossaryentry{IE802.11}{
  name={IE802.11},
  description={IEEE 802.11, commonly known as Wi-Fi, is a set of standards for wireless local area networks (WLANs)}
}

\newglossaryentry{IE802.15}{
  name={IE802.15},
  description={IEEE 802.15 is a family of standards for wireless personal area networks (WPANs)}
}

\newglossaryentry{IE802.15.6}{
  name={IE802.15.6},
  description={IEEE 802.15.6 is a standard for wireless body area networks (WBANs), defining protocols and specifications for communication between wearable and implantable medical devices}
}

\newglossaryentry{RTS}{
  name={RTS},
  description={RTS/CTS (Request to Send/Clear to Send) is a control mechanism used in communication protocols, particularly in wireless networks, to avoid data collision by coordinating the transmission of data between sender and receiver}
}

\newglossaryentry{gRPC}{
  name={gRPC},
  description={gRPC is an open-source remote procedure call (RPC) framework developed by Google that enables efficient and reliable communication between distributed systems}
}

\newglossaryentry{WiMAX}{
  name={WiMAX},
  description={WiMAX (Worldwide Interoperability for Microwave Access) is a wireless broadband technology that provides high-speed data transmission over long distances}
}

\newglossaryentry{LTE-A}{
  name={LTE-A},
  description={LTE-A (Long-Term Evolution Advanced) is a standard for wireless broadband communication, offering higher data rates and improved performance compared to previous generations of LTE technology}
}

\newglossaryentry{OpenFlow}{
  name={OpenFlow},
  description={OpenFlow is a communication protocol used in software-defined networking (SDN) to manage and control network devices}
}

\newglossaryentry{LoRa}{
  name={LoRa},
  description={LoRa (Long Range) is a wireless communication technology designed for long-range, low-power IoT applications}
}

\newglossaryentry{OPC}{
  name={OPC},
  description={Open Platform Communication (OPC) is an open-source platform for communication and collaboration, providing tools and services for sharing information, coordinating activities, and building communities across distributed environments}
}

\newglossaryentry{Implementation}{
    name={Implementation},
    description={Testing and assessing the actual system in a real-world setting}
}
\newglossaryentry{Simulation}{
    name={Simulation},
    description={Using computer models to replicate system behavior under different conditions}
}
\newglossaryentry{Testbed}{
    name={Testbed},
    description={Deploying a controlled environment for systematic system testing}
}
\newglossaryentry{FormalVerification}{
    name={FormalVerification},
    description={Employing mathematical methods to rigorously verify system correctness and adherence to requirements such as model checking}
}

\newglossaryentry{HLG}{
    name={HLG},
    description={Hyperledger blockchains are decentralized, permissioned distributed ledgers that provide a framework for developing blockchain-based applications. HLG refers collectively to Hyperledger Fabric (HLF) and Hyperledger Sawtooth}
}
\newglossaryentry{Ethereum}{
    name={Ethereum},
    description={Ethereum is an open-source, decentralized blockchain platform that enables the creation and deployment of smart contracts and decentralized applications (DApps). The Ethereum category encompasses both Ethereum and RapidChain which is a blockchain network forked from Ethereum}
}
\newglossaryentry{Bitcoin}{
    name={Bitcoin},
    description={Bitcoin is a decentralized digital currency that enables peer-to-peer transactions without the need for a central authority or intermediary}
}

\newglossaryentry{Monero}{
    name={Monero},
    description={Monero is a cryptocurrency that leverages blockchain technology along with privacy-enhancing features to obscure transactions, ensuring anonymity}
}

\newglossaryentry{RecordChain}{
    name={RecordChain},
    description={A blockchain architecture to handle big data among distributed edge nodes}
}

\newglossaryentry{MicroChain}{
    name={MicroChain},
    description={MicroChain is a blockchain scalability solution designed to improve the throughput and performance of decentralized applications (DApps) by implementing a network of interconnected sub-chains}
}

\newglossaryentry{Cosmos}{
    name={Cosmos},
    description={Cosmos is a decentralized network of independent blockchains that are interoperable and can communicate with each other, enabling the exchange of assets and data across different blockchain platforms}
}

\newglossaryentry{FISCO}{
    name={FISCO},
    description={A safe and controllable enterprise-level financial consortium blockchain platform}
}

\newglossaryentry{Tangle}{
    name={Tangle},
    description={Tangle is a data structure used in the IOTA cryptocurrency platform, designed to facilitate feeless transactions and enable scalability by using a directed acyclic graph (DAG) instead of a traditional blockchain}
}
\newglossaryentry{TrustChain}{
    name={TrustChain},
    description={TrustChain is a blockchain-based system designed to track and verify the provenance of goods and ensure their authenticity throughout the supply chain}
}

\newglossaryentry{IOTA}{
    name={IOTA},
    description={IOTA is a distributed ledger technology specifically designed for the Internet of Things (IoT) ecosystem}
}

\newglossaryentry{MultiChain}{
    name={MultiChain},
    description={MultiChain is a platform that enables the creation and deployment of private blockchains. It is designed to facilitate the development of custom blockchain solutions tailored to specific business requirements}
}

\newglossaryentry{NaiveChain}{
    name={NaiveChain},
    description={Naivchain implements basic features required for a functioning blockchain and is primarily designed for demonstration and educational purposes}
}

\newglossaryentry{EOSChain}{
    name={EOSChain},
    description={EOS/ EOS.IO is a blockchain platform to develop industrial-scale decentralized applications}
}

\newglossaryentry{LiTiChain}{
    name={LiTiChain},
    description={Litichain is a blockchain framework characterized by finite-lifetime blocks, tailored specifically for applications in the Internet of Things (IoT) and edge computing domains}
}

\newglossaryentry{XuperChain}{
    name={XuperChain},
    description={XuperChain is a blockchain platform developed by Baidu, designed to provide a high-performance, scalable, and secure infrastructure for decentralized applications (dApps) and digital asset management}
}

\newglossaryentry{VeChain}{
    name={VeChain},
    description={An enterprise-grade blockchain platform at the forefront of the sustainability revolution, offering a highly scalable smart contract platform designed to facilitate low-carbon solutions}
}

\newglossaryentry{ProofOfWork}{
    name={ProofOfWork},
    description={Proof of work is a consensus algorithm used by Bitcoin and other blockchains to ensure blocks are only regarded as valid if they require a certain amount of computational power to produce. This category also includes Time-oriented Proof of Work}
}
\newglossaryentry{BFT}{
    name={BFT},
    description={Byzantine Fault Tolerance is a consensus mechanism that allows to reach agreement among distributed nodes even if some of them are faulty or malicious}
}
\newglossaryentry{ABFT}{
    name={ABFT},
    description={Asynchronous byzantine fault tolerance (ABFT) is a type of Byzantine fault tolerant consensus algorithms, which allow for honest nodes of a network to guarantee to agree on the timing and order of a set of transactions fairly and securely}
}
\newglossaryentry{FABPaxos}{
    name={FABPaxos},
    description={Fast Byzantine Paxos is a fast consensus algorithm that is a variant of classic Paxos}
}
\newglossaryentry{RAFT}{
    name={RAFT},
    description={It has the same function as Paxos, but compared to Paxos, it is easier to understand and easier to apply to actual systems}
}
\newglossaryentry{RPCA}{
    name={RPCA},
    description={Ripple Protocol Consensus Algorithm (RPCA) checks the consensus about the ledger with the connected nodes every few seconds}
}
\newglossaryentry{ProofOfStake}{
    name={ProofOfStake},
    description={Proof-of-stake (PoS) protocols are a class of consensus mechanisms for blockchains that work by selecting validators in proportion to their quantity of holdings in the associated cryptocurrency. It includes Proof of Coin, Delegated Proof of Stake, and Delegated Consensus. This category also includes Proof Of Coin and delegated Proof Of Stake}
}
\newglossaryentry{ProofOfAuthority}{
    name={ProofOfAuthority},
    description={The proof of authority (PoA) consensus is fundamentally an improved PoS consensus that controls identity as the system of stake rather than token staking}
}
\newglossaryentry{ProofOfElapsedTime}{
    name={ProofOfElapsedTime},
    description={Proof of Elapsed Time (PoET) is a Nakamoto-style consensus algorithm where proof of work is replaced by a wait time randomly generated by a trusted execution environment (TEE)}
}
\newglossaryentry{CollectiveSign}{
    name={CollectiveSign},
    description={A consensus protocol based on signatures that relies on digital signatures to validate and authenticate transactions}
}

\newglossaryentry{ProofOfSpace}{
    name={ProofOfSpace},
    description={Proof-ofSpace (PoS) is used for mining devices in the network using their available space of hard drive for deciding mining rights and transaction validation}
}

\newglossaryentry{ProofOfPUF}{
    name={ProofOfPUF},
    description={Proof-of-PUF (PoP), also known as Proof-of-Physical-Unclonable-Function, is a consensus mechanism that leverages physical unclonable functions (PUFs) to achieve consensus in blockchain networks}
}
\newglossaryentry{ProofOfCredit}{
    name={ProofOfCredit},
    description={Proof of Credit (PoC) calculates participants' credit based on their contributions and restricts low-credit users from consensus, ensuring network integrity}
}
\newglossaryentry{ProofOfQDB}{
    name={ProofOfQDB},
    description={A lightweight feeless consensus algorithm named Poof of Quality of Service based DAGs-to-Blockchain (PoQDB)}
}
\newglossaryentry{VoteConsensus}{
    name={VoteConsensus},
    description={Voting-based consensus mechanisms achieve consensus on transactions and key network decisions by counting the number of votes cast by nodes on the network}
}

\newglossaryentry{ProofOfReputation}{
    name={ProofOfReputation},
    description={A consensus mechanism where the influence or voting power of participants is determined by their reputation or past behavior within the network}
}

\newglossaryentry{ProofOfCollaboration}{
    name={ProofOfCollaboration},
    description={A consensus mechanism in which participants must actively cooperate or work together to validate transactions and achieve consensus}
}

\newglossaryentry{ProofOfTrust}{
    name={ProofOfTrust},
    description={A consensus mechanism that relies on participants' trustworthiness or reputation within the network to validate transactions and achieve consensus}
}

\newglossaryentry{ProofOfExistence}{
    name={ProofOfExistence},
    description={A consensus mechanism that provides cryptographic proof that a particular document or transaction exists at a certain point in time without revealing the actual content}
}

\newglossaryentry{ProofOfLearning}{
    name={ProofOfLearning},
    description={A consensus mechanism where participants must demonstrate knowledge or learning capabilities to contribute to the validation process and achieve consensus}
}

\newglossaryentry{ProofOfEfficiency}{
    name={ProofOfEfficiency},
    description={A consensus mechanism that rewards participants based on their efficiency in processing transactions or performing computational tasks within the network}
}

\newglossaryentry{Tendermint}{
    name={Tendermint},
    description={A consensus algorithm that employs a Byzantine Fault Tolerance (BFT) approach and relies on a set of validators to achieve consensus through a voting process}
}

\newglossaryentry{ProofOfService}{
    name={ProofOfService},
    description={A consensus mechanism where participants provide evidence of providing a specific service or performing a task to validate transactions and achieve consensus}
}

\newglossaryentry{ProofOfUsefulWork}{
    name={ProofOfUsefulWork},
    description={A consensus mechanism that requires participants to perform useful computational work or solve specific problems to validate transactions and achieve consensus}
}

\newglossaryentry{Solo}{
    name={Solo},
    description={A consensus mechanism where a single node or entity validates transactions and creates new blocks without the need for validation by other participants}
}
\makeglossaries

\begin{document}

\title{Blockchain and Edge Computing Nexus: A Large-scale Systematic Literature Review}

\author{Zeinab Nezami, Zhuolun Li, Chuhao Qin, Fatemeh Banaie, Rabiya Khalid, \\ Evangelos Pournaras
\IEEEcompsocitemizethanks{\IEEEcompsocthanksitem ZN conducted the work with the School of Computer Science and is currently with the School of Electronic and Electrical Engineering, University of Leeds, Leeds, UK. FH conducted the work with the School of Computer Science, University of Leeds, Leeds and is currently with the Department of Computer Science, University of Huddersfield, Huddersfield, UK. ZL, CQ, RK and EP are with the School of Computer Science, University of Leeds, Leeds, UK.
\protect\\
Emails: \{z.nezami,sczl,sccq,scsrkh,e.pournaras\}@leeds.ac.uk,\\
f.banaieheravan@hud.ac.uk
}
}

\maketitle

\begin{abstract}

Blockchain and edge computing are two instrumental paradigms of decentralized computation, driving key advancements in Smart Cities applications such as supply chain, energy and mobility. Despite their unprecedented impact on society, they remain significantly fragmented as technologies and research areas, while they share fundamental principles of distributed systems and domains of applicability. This paper introduces a novel and large-scale systematic literature review on the nexus of blockchain and edge computing with the aim to unravel a new understanding of how the interfacing of the two computing paradigms can boost innovation to provide solutions to timely but also long-standing research challenges. By collecting almost 6000 papers from 3 databases and putting under scrutiny almost 1000 papers, we build a novel taxonomy and classification consisting of 22 features with 287 attributes that we study using quantitative and machine learning methods. They cover a broad spectrum of technological, design, epistemological and sustainability aspects. Results reveal 4 distinguishing patterns of interplay between blockchain and edge computing with key determinants the public (permissionless) vs. private (permissioned) design, technology and proof of concepts. They also demonstrate the prevalence of blockchain-assisted edge computing for improving privacy and security, in particular for mobile computing applications. 
\end{abstract}


\section{Introduction}\label{sec:intro}
\IEEEPARstart{T}{he} imminent surge in Internet of Things (IoT) devices, forecasted to surpass 55.7 billion by 2025 and generate up to 79.4 zettabytes of data~\cite{rydning2018digitization}. Yet, a mere fraction, less than 0.5\%, undergoes analysis and utilization\footnote{Myler L. Better Data Quality Equals Higher Marketing ROI. Available at \url{https://www.forbes.com/sites/larrymyler/2017/07/11/better-data-quality-equals-higher-marketing-roi/?sh=53fe05867b68}, accessed June 2025.}. This phenomenon arises due to various factors, such as constrained budgets for data analysis, deficiencies in data integration tools, manual data entry and collection procedures, as well as the use of numerous standalone analytical tools, among others.
On the other hand, an IBM study reveals the staggering costs associated with data breaches, with an average global annual cost estimated at \$4.24 million, and lost business accounting for \$1.59 million or 38\% of the total cost. 

The emergence of edge and fog computing (referred to as edge computing in this paper) becomes a transformative paradigm, positioning computation and data storage closer to application and end users. Edge computing addresses the limitations of traditional cloud-based architectures: as IoT-generated data processing requires for low-latency and high-bandwidth applications escalate, the decentralized approach of edge computing enhances efficiency by enabling computations at the source, thereby reducing latency and associated costs. 

In 2023, the global edge computing market amounted to \$15.96 billion, with projections indicating robust growth, surging to around \$216.76 billion by 2032, yielding a substantial compound annual growth rate (CAGR) of 36.6\%\footnote{Edge Computing Technology Market Size. Available at \url{https://www.fortunebusinessinsights.com/edge-computing-market-103760}, last accessed June 2025.}. Simultaneously, blockchain technology garners attention for its potential to revolutionize industries by enhancing transparency, security, and trust in decentralized systems. Originally underpinning cryptocurrencies, blockchain has diversified into various sectors, including IoT, supply chain management, and healthcare, leveraging its immutable, decentralized, and cryptographically secure principles. The global blockchain technology market reached \$17.57 billion in 2023, projected to skyrocket to \$825.93 billion by 2032, with an impressive CAGR of 52.8\%\footnote{Blockchain Technology Market Size. Available at \url{https://www.fortunebusinessinsights.com/industry-reports/blockchain-market-100072}, last accessed May 2024.}.

The convergence of edge computing and blockchain offers a promising avenue to tackle challenges inherent in centralized systems. By melding decentralized architectures of blockchain with the proximity and agility of edge computing, applications with real-time data processing, secure transactions, improved privacy and security are gaining momentum. Despite the growing interest in integrating edge computing and blockchain, a comprehensive understanding of the existing research landscape is crucial and missing. This systematic literature review (SLR) addresses this gap by analyzing and synthesizing literature, offering insights into key themes, methodologies, challenges, and opportunities.

DePIN, or Decentralized Physical Infrastructure Networks, utilizes blockchain technology and token incentives to support real-world infrastructures spanning to transportation, energy, and computing domains~\cite{Ballandies2023,Lin2024}. This innovative approach capitalizes on blockchain features to meet the rising need for distributed physical infrastructures. By enabling widespread participation from individuals and organizations, DePINs offer financial rewards and ownership opportunities, heralding a transformative shift in infrastructure development and governance\footnote{DePIN: Unlocking New Opportunities For Scalable Cloud Infrastructure. Available at \url{https://www.forbes.com/sites/forbestechcouncil/2024/05/09/depin-unlocking-new-opportunities-for-scalable-cloud-infrastructure/}, accessed June 2025}.

In this paper, we delve into the nexus of edge computing and blockchain technologies, aiming to provide insights into the current state, emerging trends, and research gaps. By synthesizing peer-reviewed articles and conference papers, we lay the groundwork for future research, inform industry practices, and guide policymakers in leveraging these transformative technologies by answering the following research questions:

{\textbf{Research Question 1}}: How the two blockchain and edge computing interact and which criteria distinguish these interactions between them?

{\textbf{Research Question 2}}: How does the blockchain and edge computing nexus co-evolve? In particular, we explore: (i) What are the increasingly important problems the blockchain and edge computing nexus are addressing? How do the two computing paradigm solve these problems? (ii) What are the choices of blockchain platforms, consensus, and reward mechanisms for interfacing with edge computing systems?

{\textbf{Research Question 3}}: What are the key study patterns that emerge in research on the nexus of edge computing and blockchain?

To answer the research questions, this paper makes the following contributions:

\begin{enumerate}
    \item A taxonomy for edge computing and blockchain systems that systematically reviews a collection of 22 descriptive and qualitative characteristics, encompassing a comprehensive range of possible values for each characteristic.
    \item A rigorous classification of 921 blockchain and edge computing systems backed by an extensive literature review.
    \item A multi-dimensional analysis of the blockchain and edge computing nexus using machine learning methods to derive an in-depth understanding of how the two computer paradigms interacts and co-evolve.
    \item New insights derived from the associations of characteristics, temporal analysis and the four key patterns identified. 
    \item Open (meta)data of the conducted systematic literature review made available to encourage further research and analysis\footnote{\url{https://doi.org/10.5281/zenodo.15615365} last accessed June 2025}
\end{enumerate}

The rest of this paper is structured as follows: Section~\ref{sec:background} provides an overview of the background and shows how this study addresses the existing gaps in the literature. Following that, Section~\ref{sec:taxonomy} introduces a new taxonomy and Section~\ref{sec:method} outlines the paper selection process and meta-analysis methods. The findings, encompassing descriptive statistics and insights derived from machine learning techniques, are presented in Section~\ref{sec:results}, followed by a comprehensive discussion in Section~\ref{sec:disc}. Lastly, Section~\ref{sec:conc} offers concluding remarks and outlines avenues for future research.

\section{Background and Literature Review}\label{sec:background}
This section delineates the scope of the review within the interface of edge computing and blockchain, breaks down the nexus of these two computing diagram into three perspectives
for the study of their interactions.

\subsection{Blockchain and Edge Computing}\label{sec:bc&ec}
Edge computing~\cite{shi2016edge} is a type of distributed system that stores and processes data using resources close to the location of users, rather than relying on a centralized data-processing warehouse~\cite{Nezami2021}. This computing paradigm emerged from the need to reduce latency and bandwidth use, enabling real-time data processing and decision-making, which is crucial for applications such as autonomous vehicles, smart cities, and IoT~\cite{Nezami2025}.

Blockchain~\cite{swan2015blockchain,ballandies2022decrypting}, on the other hand, is a decentralized ledger technology that ensures secure, transparent, and tamper-proof record-keeping. Originally developed to support cryptocurrencies such as Bitcoin~\cite{nakamoto2008bitcoin}, blockchain has expanded to various applications due to its ability to provide immutable records and fault tolerance consensus.

With different advantages, the nexus of edge computing and blockchain has the potential to innovate for different applications. For example, healthcare is an application that requires fast data processing and high-end security solutions~\cite{abdellatif2021medge}. In such systems, edge computing is able to reduce latency and use bandwidth more effectively, while blockchain enhances security and trust in decentralized systems. Together, they create robust solutions capable of handling the increasing demands for real-time data processing and secure transactions. This synergy enables the creation of innovative solutions that enhance efficiency, transparency, and trust in various domains, from Smart Cities and digital democracy~\cite{Pournaras2018,Pournaras2020,Li2024}, supply chain management~\cite{rizwan2022internet} to financial services~\cite{wang2022supply} and transportation logistics~\cite{mei2022blockchain,Qin2023}. 

\subsection{Perspectives of Blockchain and Edge Computing Interactions}\label{sec:perspective}
To further understand how these two computing paradigms interact, this work, for the first time, categorizes and investigates their nexus in three distinct perspectives: (i) blockchain as a solution for edge computing challenges (Perspective 1), (ii) edge computing as a solution for blockchain challenges (Perspective 2), and (iii) the synergistic integration of both computing paradigms (Perspective 3).

\subsubsection{Perspective 1}

This perspective features blockchain-assisted edge computing systems. Blockchain provides features such as immutability~\cite{ma2022toward,samy2022secure}, transparency~\cite{jo2018hybrid,rawat2019fusion}, and decentralized consensus~\cite{du2022accelerating,rastoceanu2022blockchain,al2021proof}, which can enhance the security~\cite{shukla2021anomaly,huang2023multi}, trustworthiness~\cite{wu2022bring,bhawana2022best}, and integrity of data and transactions in edge scenarios~\cite{shukla2021anomaly,ye2021blockchain,kertesz2022block}. Researchers explore the integration of blockchain into edge computing architectures to facilitate secure data sharing~\cite{samy2022secure,ma2022toward,zhang2022blockchain}, provenance tracking~\cite{sharma2017software,lautert2020fog}, and verifiable execution of edge tasks~\cite{zhang2022blockchain,iftikhar2020efficient}. 

In addition, blockchain-enabled edge computing frameworks empower edge devices to engage in decentralized networks, fostering autonomous decision-making, self-governance, and secure data monetization. This approach expands the capabilities of edge computing, ensuring data integrity and privacy across diverse domains including healthcare~\cite{guo2022hybrid,kumar2022securing}, energy management~\cite{alkhiari2022blockchain,wang2021blockchain}, and industrial IoT (IIoT)~\cite{kumar2022blockchain,gupta2022blockchain,kumar2022securing}.

\subsubsection{Perspective 2} 

This perspective captures edge-assisted blockchain systems. Researchers and practitioners explore methods to leverage edge computing infrastructure to enhance the performance~\cite{zuo2021delay,li2021multi,xu2021deep}, scalability~\cite{qu2021enable,al2021permissioned}, and efficiency~\cite{abdellatif2021medge,chen2021distributed} of blockchain networks. By distributing blockchain nodes across edge devices and gateways, organizations aim to reduce blockchain network latency~\cite{ni2021fast,yang2021dcab}, improve data throughput~\cite{zuo2021computation,yang2021dcab,al2021permissioned}, and optimize resource utilization~\cite{huang2021resource,zhang2022truthful,zhang2022optimal}. Edge solutions for blockchain also encompass the development of lightweight consensus mechanisms~\cite{wadhwa2022energy}, efficient data/resource management strategy~\cite{Chen8720039Cooperative,Zhao8725560Coalition}, and edge-based smart contract execution engines~\cite{wang2022smart}. These advancements hold promise for enabling decentralized applications (DApps) that operate seamlessly in edge environments, catering to use cases such as IoT-driven supply-chain management~\cite{wadhwa2022energy,rizwan2022internet}, ITS~\cite{Hammedi9729800Toward}, smart cities~\cite{Masuduzzaman9728721UxV}, and smart energy systems~\cite{xu2021deep}.

\subsubsection{Perspective 3} 

This perspective encompasses the combined synergy of edge computing and blockchain paradigms. As a combination of Perspective 1 and 2, this perspective focus on the proposals that make use of the characteristics of both edge-assisted blockchain and blockchain-assisted edge computing systems~\cite{Ye9521181Blockchain,Lee8993056Secure,xu2022mudfl}. This integrated approach enables edge devices to securely interact with blockchain networks, execute smart contracts, and participate in consensus mechanisms while benefiting from localized data processing and real-time analytics~\cite{Yu2019Building,Lee8993056Secure,SALUJA2022100521}. Applications range from distributed energy trading and decentralized finance (DeFi)~\cite{zetzsche2020decentralized} to secure edge AI inference and federated learning~\cite{JAYAKUMAR20221795,xu2022mudfl}.

This systematic literature review analyzes existing research and developments across the three perspectives, representing the core interactions and potential enhancements in the synergy of edge computing and blockchain ecosystems. Distinguishing these three perspectives also provides a better understanding of the relationship between the two computing paradigm in their synergy, which is not yet comprehensively explored in related work.

\subsection{Related Work}\label{sec:relatedwork}

The majority of existing studies for blockchain and edge computing have primarily outlined the security and privacy challenges of edge computing, while suggesting blockchain as a viable solution~\cite{alzoubi2022blockchain, alhumam2023cyber, khan2023security, jiang2022distributed, gadekallu2021blockchain, gupta2021taxonomy, mendiboure2020survey}, which is described in Perspective 1 defined above. Nonetheless, these studies often overlook the opportunities that edge computing can bring into the blockchain computing paradigm and the reciprocal relationship between the two paradigms (Perspective 2 and 3). Table~\ref{tab:lit-rev} provides a summary of key surveys, highlighting their distinct research focus. 

While existing research offers valuable insights into specific applications of blockchain and edge computing, it often examines these technologies in isolation, focusing narrowly on individual aspects such as security, resource management, or data handling~\cite{jiang2022distributed,alzoubi2022blockchain,gadekallu2021blockchain,gupta2021taxonomy,mendiboure2020survey,tariq2019security,fernandez2019towards,alam2022blockchain,baranwal2023blockchain,liu2021towards,yeow2017decentralized,george2019light,xue2023integration,baniata2020survey,yang2019integrated,archana2024literature,rajesh2023review,wang2023synergy, bhat2020edge}. However, these studies lack comprehensive evaluation criteria and systematic reporting methods for their findings. Without a holistic analysis, it is challenging to understand the broader implications, synergies, and evolving dynamics of combining blockchain with edge computing and thereby uncover trends and opportunities.

\begin{table*}[htbp]
    \centering
    \caption{Existing literature reviews from 2017 to 2024 are on smaller scales and limited perspectives }
    \label{tab:lit-rev}
    \tiny
    \begin{tabular}{p{0.8cm}p{0.4cm}p{2.2cm}p{1.2cm}p{1.3cm}p{1cm}p{1cm}p{2cm}}
    \toprule
       \multicolumn{8}{c}{Non-Systematic Literature Reviews}\\
       \midrule
       \multicolumn{3}{c}{Studies}&\multicolumn{3}{c}{Areas of focus}&\multicolumn{2}{c}{Perspectives}\\
       \midrule
       \multicolumn{3}{l}{\cite{archana2024literature,alzoubi2022blockchain,tariq2019security}} & \multicolumn{3}{l}{Security and privacy challenges} & \multicolumn{2}{l}{Perspective 1} \\
        \multicolumn{3}{l}{\cite{gupta2021taxonomy,jiang2022distributed,mendiboure2020survey}} & \multicolumn{3}{l}{Application: Intelligent Transport System} & \multicolumn{2}{l}{Perspective 1}\\

        \multicolumn{3}{l}{\cite{gadekallu2021blockchain,baniata2020survey}} & \multicolumn{3}{l}{Applications and Challenges} & \multicolumn{2}{l}{Perspective 1} \\

        \multicolumn{3}{l}{\cite{xue2023integration}} & \multicolumn{3}{l}{Application: IoT} & \multicolumn{2}{l}{Perspective 1} \\

        \multicolumn{3}{l}{\cite{rajesh2023review,liu2021towards}} & \multicolumn{3}{l}{Application: Smart manufacturing} & \multicolumn{2}{l}{Perspective 1} \\

        \multicolumn{3}{l}{\cite{bhat2020edge}} & \multicolumn{3}{l}{Frameworks and challenges} & \multicolumn{2}{l}{Perspective 1} \\
        
        \multicolumn{3}{l}{\cite{yang2019integrated,wang2023synergy}} & \multicolumn{3}{l}{Frameworks and challenges} & \multicolumn{2}{l}{Perspective 1, 2 ,3} \\
        
        
       \multicolumn{3}{l}{\cite{alam2022blockchain}} & \multicolumn{3}{l}{Application: healthcare} & \multicolumn{2}{l}{Perspective 1}\\

       \multicolumn{3}{l}{\cite{baranwal2023blockchain}} & \multicolumn{3}{l}{Resource allocation} & \multicolumn{2}{l}{Perspective 1}\\

       \multicolumn{3}{l}{\cite{yeow2017decentralized}} & \multicolumn{3}{l}{Blockchain design choices} & \multicolumn{2}{l}{Perspective 1}\\

       \multicolumn{3}{l}{\cite{george2019light}} & \multicolumn{3}{l}{Cryptography solutions} & \multicolumn{2}{l}{Perspective 2}\\
        \toprule
        \multicolumn{8}{c}{Systematic Literature Reviews}\\
        \midrule
        Study&year&Area of focus&Perspective&Literature time range&Sample size (size after filtering)&Num. features (Num. attributes)&Type of analysis\\
        \midrule
       \cite{alzoubi2023blockchain}&2023&Applications \& challenges&Perspective 1&2016-Aug 2021&557 (144)&9 (38)&Qualitative\\

       \cite{alhumam2023cyber}&2023&Security&Perspective 1&Not mentioned&1833 (20)&4&Qualitative\\

       \cite{khan2023security}&2023&Security&Perspective 1&2019-2022&150 (48)&3&Qualitative\\ 

       \cite{alzoubi2022systematic}&2022&Purposes of integration&Perspective 1&2016-Apr 2022&517 (181)&7 (28)&Qualitative\\

       This work&2024&Applications, problems and study patterns&Perspective 1,2,3&2015-Oct 2022&5903 (921)&22 (287)&Qualitative and Quantitative\\
       \bottomrule
    \end{tabular}
\end{table*}

To the best of our knowledge, recent publicly available studies have not conducted a systematic review on a large scale of papers (with less than 200 papers after filtering) with a comprehensive quantitative analysis of the existing literature regarding the synergies of blockchain and edge computing paradigms. The bottom of Table~\ref{tab:lit-rev} represents the related systematic studies and their characteristics. Alzoubi and Aljaafreh~\cite{alzoubi2023blockchain} study various applications of applying blockchain to edge computing systems, while Alhumam\textit{et al.}~\cite{alhumam2023cyber} and Khan\textit{et al.}~\cite{khan2023security} primarily focus on the security solutions provided by blockchain for edge computing. On the other hand, Alzoubi\textit{et al.}~\cite{alzoubi2022systematic} explore the seven key motivations behind blockchain-assisted edge computing, including security, privacy, access control, trust management, data management, scalability, and performance. Contrary to these studies that solely explore one specific aspect of blockchain-assisted edge computing systems, this paper covers 22 key aspects of recent research on blockchain and edge computing synergy.

This paper goes beyond the scope of individual applications and comprehensively reviews the existing work in the field from several perspectives (e.g., problem, contribution, application, security, privacy, evaluation, incentives, technology, and so on) rigorously and systematically. While a qualitative descriptive analysis is the only insight provided by the literature~\cite{alzoubi2022systematic,alhumam2023cyber,khan2023security,alzoubi2023blockchain}, this work, for the first time, applies quantitative machine learning techniques on a large-scale dataset derived from this systematic literature review to explore the co-evolution, interaction, and study patterns of the nexus of blockchain and edge computing.

\section{Taxonomy}\label{sec:taxonomy}

The taxonomy introduced in this study is constructed using the methodology proposed by Nickerson~\etal{}~\cite{nickerson2013method}, which has also proven earlier effective to use in the domain of computing and blockchain~\cite{ballandies2022decrypting}. Its primary objective is to facilitate a comprehensive classification of blockchain-edge computing systems, aiming to quantitatively extract key design choices within these systems. The study dimensions cover research aspects, system design aspects, and broader socio-economic aspects.

Under research aspects, we focus on identifying and categorizing the specific problems addressed in the literature, examine the practical applications of these paradigms across various industries, and exploring the scientific and technical contributions. The evaluation methods used in studies are also assessed to ensure the reliability and validity of findings, and the metrics for success and performance measurement are identified.

For system design aspects, we study the integration of artificial intelligence (AI) methods, security and privacy advantages offered by blockchain, examine resource allocation strategies for efficiency and optimization, and assess the underlying technologies and communication protocols used in these systems. Broader socio-economic aspects include examining sustainability implications, evaluating the Technology Readiness Level (TRL) to understand the maturity and real-world application potential, and investigating blockchain operational and economic models via dimensions such as blockchain, permission, type, consensus mechanism, and reward.

In addition to studying various aspects within the converging realms of blockchain and edge computing, this study also aims to highlight the open science initiatives surrounding these converging fields. Two specific dimensions in our review focus on listing the open-source software and datasets contributed by the studied papers. 

As a result, each article within the final sample is classified and coded using 22 design dimensions formulated for blockchain and edge computing, as listed in Table~\ref{tab:tax-glos}. These dimensions are derived through a thorough examination and analysis of the domain~\cite{farahani2021convergence,gadekallu2021blockchain,yang2019integrated,xue2023integration,ballandies2022decrypting}, followed by iterative refinement through in-depth reviews for attribute extraction and coding updates. The complete dataset, containing the fully coded set of studies along with the code used for subsequent analyses and visualizations, is publicly available\footnote{A vailable at \url{https://github.com/TDI-Lab/Edge-Blockchain-Nexus} (last accessed: December 2024)}. A detailed outline of the dimensions and their constituent attributes are presented below, along with citations from the most recent published studies we reviewed.

\begin{table*}[htbp]
    \centering
    \caption{The 22 dimensions consisting of the proposed taxonomy.}
    \label{tab:tax-glos}
    \tiny
    \begin{tabularx}{.96\textwidth}{p{1.2cm}p{13cm}}
        \toprule
        Dimension&Attributes\\
        \midrule
        Scope&Non-Methodological (0), Methodological (1)\\
        
        Application&\gls{VideoStreaming}, \gls{Healthcare}, \gls{SmartEnergy}, \gls{IIoT}, \gls{IoT}, \gls{SmartCity}, \gls{WeatherForecast}, \gls{Drones}, \gls{CrowdSourcing}, \gls{MobileComputing}, \gls{5G}, \gls{6G}, \gls{SmartTransport}, \gls{SupplyChain}, \gls{ECommerce}, \gls{SmartFarm}, \gls{SmartWater}, \gls{Surveillance}, \gls{SmartCameras}, \gls{SmartHome}\\
        
        Problem&\gls{DataSharing}, \gls{DataTrading}, \gls{Performance}, \gls{Trust}, \gls{Privacy}, \gls{Security}, \gls{Transparency}, \gls{problemReliability}, \gls{problemScalability}, \gls{Incentivization}, \gls{Programmability}, \gls{Experimentation}, \gls{DataProvenance}, \gls{DataRecovery}\\
        
        Contribution&\gls{Model}, \gls{Framework}, \gls{Platform}, \gls{Architecture}, \gls{Methodology}, \gls{Algorithm}, \gls{Protocol}, \gls{SoftwareArtifact}, \gls{Service}, \gls{Mechanism}\\
        
        AI method&\gls{Reinforcement}, \gls{Federated}, \gls{Unsupervised}, \gls{Supervised}, \gls{Heuristic}, \gls{Metaheuristic}, \gls{aiMatching}, \gls{Reasoning}\\
        
        Security&\gls{Authentication}, \gls{Encryption}, \gls{AccessControl}, \gls{Verifiability}, \gls{TrustManagement}, \gls{Credibility}, \gls{Integrity}, \gls{SecurityAvailability}, \gls{Immutability}, \gls{Confidentiality}, \gls{Authorization}, \gls{NonRepudiation}\\
        
        Privacy&\gls{Anonymity}, \gls{DiffPrivacy}, \gls{OPRF}, \gls{Encryption}, \gls{PSI}, \gls{HomoEncryption}, \gls{RingSignature}, \gls{ZeroKnowlProof}, \gls{Pseudonymity}, \gls{BlindSignature}, \gls{SecretSharing}, \gls{AutoEncoder}, \gls{HybridIdentity}\\
        
        Allocation&\gls{Trading}, \gls{Offloading}, \gls{Balancing}, \gls{Placement}, \gls{scaling}, \gls{Scheduling}, \gls{allocMatching}\\
        
        Metric&\gls{CompDelay}, \gls{Throughput}, \gls{SecurityVulner}, \gls{Latency}, \gls{GasConsumption}, \gls{UtilSocialWelfare}, \gls{ConvergeSpeed}, \gls{CommCost}, \gls{CompCost}, \gls{Accuracy}, \gls{PowerConsumption}, \gls{EnergyConsumption}, \gls{CommUtil}, \gls{CompUtil}, \gls{StorageUtil}, \gls{MetricReliability}, \gls{MetricAvailability}, \gls{Precision}, \gls{Recall}, \gls{F1Score}, \gls{StorageCost}, \gls{FailureCost}, \gls{metricPrivacy}, \gls{LinesOfCode}, \gls{Reputation}, \gls{WinningRate}, \gls{metricScalability}, \gls{Jitter}\\
        
        Technology& \gls{IPFS}, \gls{RFID}, \gls{CellularNet}, \gls{MEC}, \gls{Docker}, \gls{InfluxDB}, \gls{Tech5G}, \gls{Tech6G}, \gls{WiFi}, \gls{Bluetooth}, \gls{Zigbee}, \gls{ANT}, \gls{NFC}, \gls{SDN}, \gls{NFV}, \gls{LightNet}, \gls{ETHTestnet}, \gls{Redis}, \gls{OpenStack}, \gls{CouchDB}, \gls{Kafka}, \gls{BigchainDB}, \gls{GoETH}, \gls{RaspberryPi}, \gls{Nodejs}, \gls{Mininet}, \gls{REST}, \gls{Arduino}, \gls{NodeRED}, \gls{iFogSim}, \gls{Brain4Net}, \gls{EdisonSoC}, \gls{Wireshark}, \gls{Jolinar}, \gls{GPS}, \gls{Contiki}, \gls{OPNET}, \gls{GMapAPI}, \gls{Chainlink}, \gls{NS3}, \gls{Postman}, \gls{BlackPill}, \gls{HealthShield}, \gls{JMeter}, \gls{OpenThread}, \gls{AVISPA}, \gls{Onos}, \gls{Prometheus}, \gls{NOMA}, \gls{RoF}, \gls{SQLite}, \gls{OrbitDB}, \gls{Ganache}, \gls{OneSwarm}, \gls{RemixIDE}, \gls{JetXavNx}, \gls{CIDDS}, \gls{PostgreSQL},  \gls{OMNeT}, \gls{SQL}, \gls{MYSQL}, \gls{SUMO}, \gls{K8S}, \gls{LevelDB}, \gls{JMTSim}, \gls{FoBSim}, \gls{IntelSGX}, \gls{JetsonNano}, \gls{ZooKeeper}, \gls{Caliper}\\
        
        TRL&0, 1, 2, 3, 4, 5, 6, 7, 8, 9\\
        
        Open dataset&Link to the proposed dataset\\
        
        Open source&Link to the proposed source code\\
        
        Sustainability&\gls{NoPoverty}, \gls{ZeroHunger}, \gls{HealthWellbeing}, \gls{EducationQuality}, \gls{GenderEquality}, \gls{WaterSanitation}, \gls{Energy}, \gls{WorkEconomic}, \gls{IndustryStructure}, \gls{ReduceInequality}, \gls{SustainableCity}, \gls{ResponsibleConsProd}, \gls{ClimateAction}, \gls{LifeBelowWater}, \gls{LifeOnLand}, \gls{PeaceJustice}, \gls{GlobalPartnership}\\
        
        Communication&\gls{MQTT}, \gls{VANET}, \gls{TLS}, \gls{M2M}, \gls{CAN}, \gls{OFDM}, \gls{SoAP}, \gls{CoAP}, \gls{STEP}, \gls{IE1609.2}, \gls{IE802.11}, \gls{IE802.15}, \gls{RTS}, \gls{gRPC}, \gls{WiMAX}, \gls{LTE-A}, \gls{OpenFlow}, \gls{LoRa}, \gls{OPC}, \gls{IE802.15.6}\\
        
        Evaluation&\gls{Implementation}, \gls{Simulation}, \gls{Testbed}, \gls{FormalVerification}\\
        
        Blockchain& \gls{HLG}, \gls{Ethereum}, \gls{Bitcoin}, \gls{Monero}, \gls{RecordChain}, \gls{MicroChain}, \gls{Cosmos}, \gls{FISCO}, \gls{Tangle}, \gls{TrustChain}, \gls{IOTA}, \gls{MultiChain}, \gls{NaiveChain}, \gls{EOSChain}, \gls{LiTiChain}, \gls{XuperChain}, \gls{VeChain}, \gls{EOSChain} \\
        
        Consensus&\gls{ProofOfWork}, \gls{BFT}, \gls{ABFT}, \gls{FABPaxos}, \gls{RAFT}, \gls{RPCA}, \gls{ProofOfStake}, \gls{ProofOfAuthority}, \gls{ProofOfElapsedTime}, \gls{CollectiveSign}, \gls{ProofOfSpace}, \gls{ProofOfPUF}, \gls{ProofOfCredit}, \gls{ProofOfQDB}, \gls{VoteConsensus}, \gls{ProofOfReputation}, \gls{ProofOfCollaboration}, \gls{ProofOfTrust}, \gls{ProofOfExistence}, \gls{ProofOfLearning}, \gls{ProofOfEfficiency}, \gls{Tendermint}, \gls{ProofOfService}, \gls{ProofOfUsefulWork}, \gls{Solo}\\
        
        Permission&Permissioned (0), Permissionless (1)\\
        
        Type&Public (0), Private(1), Hybrid(2), Consortium (3)\\
        
        Chain&Off-Chain (0), On-Chain (1), Both (2)\\
        
        Reward&Monetary (0), Reputation (1), Hybrid (2), Multi-Dimension (3)\\
        \bottomrule
    \end{tabularx}
\end{table*}

\noindent \textbf{Scope} is a dimension for papers of a methodological nature or contribution (value of 1, and a value of 0 for otherwise)~\cite{masuduzzaman2022uxv,ahmed2022blockchain, nguyen2021bedgehealth}. In the domain of blockchain and edge computing, a methodological paper is one that introduces novel methodologies to assess, prototype, deploy, or evaluate the performance of systems integrating both blockchain and edge computing paradigms~\cite{lakhan2022its,xu2022certificateless,kochovski2021building}.

\noindent \textbf{Application} refers to the application scenario studied in the paper. It delves into the practical use cases and contexts where the integration of blockchain and edge computing is explored. The application scenarios involve various domains such as IoT~\cite{xu2022mudfl,wang2022dag,kochovski2021building}, smart transportation~\cite{baniata2022prifob,lakhan2022its}, smart home~\cite{bhawana2022best,qashlan2021privacy}, smart city~\cite{hati2022dewbcity, li2022zero}, smart energy~\cite{gao2021fogchain,guan2021blockchain}, smart healthcare~\cite{nguyen2021bedgehealth,baniata2022prifob}, industrial applications~\cite{hewa2022fog,rahman2022blockchain,lin2022novel}, 6G~\cite{ozdogan2022digital,abdulqadder2022sliceblock}, surveillance~\cite{datta2022bssffs,yu2022blockchain}, crowd-sourcing~\cite{hu2022trusted,pardeshi2022hash}, and more. 

\noindent \textbf{Problem} denotes the core challenges that the paper aims to tackle. Challenges might include issues related to fault-tolerance, security~\cite{wang2022dag,baniata2022prifob}, privacy~\cite{baniata2022prifob,xu2022mudfl}, performance~\cite{baniata2022prifob,xu2022mudfl,wang2022dag}, reliability~\cite{nguyen2021bedgehealth,wang2022dag}, scalability~\cite{baniata2022prifob,xu2022mudfl,wang2022dag}, data sharing~\cite{baniata2022prifob,cui2021secure}, data trading~\cite{lin2020blockchain,nam2022iot}, programmability~\cite{kochovski2021building}, and other overarching problems pertinent to the integration of blockchain and edge computing systems.

\noindent \textbf{Contribution} outlines what the paper brings to the field. Contributions could take various forms, such as introducing a new software artifact~\cite{xu2022decentralized,nam2022iot,kochovski2021building}, proposing a framework~\cite{baniata2022prifob,yuan2022jora}, developing a novel architecture~\cite{lin2022intelligent,xiao2022consortium,hewa2022fog}, introducing findings and insights, creating a new mechanism~\cite{xu2022mudfl,wang2022dag}, and other. 

\noindent \textbf{AI method} identifies the artificial intelligence, learning, or optimization techniques used in the paper. Examples include fuzzy inference~\cite{jayakumar2022design,gardas2022fuzzy}, supervised learning methods~\cite{almalki2022enabling,guo2022blockchain}, unsupervised methods~\cite{priscila2022risk,du2022accelerating}, reinforcement learning~\cite{guo2022incentive,heidari2022deep}, federated learning~\cite{xu2022mudfl,abdel2022privacy}, optimization heuristics~\cite{yuan2022jora,zhang2021bpaf}, and other. 

\noindent \textbf{Allocation} assesses the resource allocation and resource management aspects within the realms of blockchain and edge computing. The challenges explored in this context may encompass load balancing~\cite{li2022farda,tulkinbekov2022blockchain}, offloading~\cite{li2022intelligent,song2022multimedia}, trading~\cite{xu2022decentralized,nam2022iot}, matching~\cite{chen2022internet,wang2022blockchain}, and other. Papers contributing to this dimension typically delve into the optimization, allocation, or efficient utilization of resources in decentralized environments of edge computing and blockchain. Moreover, the AI method dimension offers a solution to tackle resource allocation challenges within this dimension.

\noindent \textbf{Metrics} refers to the quantitative assessments employed to evaluate the performance or efficacy of a proposed solution. Examples of metrics encompassed within this dimension include, but are not limited to: economical cost~\cite{xie2021resource,mei2022blockchain}, transaction costs/throughput~\cite{baniata2022prifob,xu2022mudfl}, latency~\cite{xu2022mudfl,wu2022bring,baniata2022prifob}, resource utilization~\cite{xu2022mudfl,jayakumar2022design,al2021proof}, energy consumption~\cite{vladyko2022distributed,saluja2022blockchain}, reliability~\cite{duhayyim2022integration,kouicem2020decentralized} and accuracy of prediction models~\cite{alkhiari2022blockchain,rahman2022blockchain}. 

\noindent \textbf{Technology} assesses the technologies utilized or proposed within the reviewed papers. Examples of such technologies include Kafka~\cite{nam2022iot,mayer2021fogchain}, SDN~\cite{latif2022sdblockedge,abdulqadder2022sliceblock}, Go-Ethereum~\cite{carvalho2021security,sylla2021blockchain}, Raspberry Pi~\cite{weerapanpisit2022decentralized,ahmed2022blockchain}, IPFS~\cite{nguyen2021bedgehealth,wang2022dag}, PostgreSQL~\cite{wang2022dag}, 5G~\cite{lei2022gbrm,hewa2022fog}, Docker~\cite{baniata2022prifob,salim2022blockchain}, Intel Software Guard Extensions (IntelSGX)~\cite{han2022blockchain,taskou2022blockchain}, and others. 

\noindent \textbf{TRL} refers to the technology readiness level of the contributed solution. TRL offers valuable insights into the advancement of technologies and their readiness for implementation or commercialization. It is a scale ranging from 1 to 9, where a higher TRL~\cite{xu2022decentralized,nam2022iot,rastoceanu2022blockchain,hou2020design,hou2020design} indicates greater maturity. The code ``0" is used to indicate non-relevance to TRL assessment.

\noindent \textbf{Open dataset} involves listing the open datasets proposed in a paper~\cite{debe2020blockchain,debe2020monetization}. 

\noindent \textbf{Open software} involves providing links to open-source software repositories proposed by a study~\cite{baniata2022prifob,meese2022bfrt,zhou2021building}. 

\noindent \textbf{Communication} assesses whether the paper studies or contributes to communication technologies or protocols. Examples include IEEE 802.15~\cite{sharmila2021edge}, RTS (Request to Send)~\cite{jiang2020intelligent}, gRPC (Google Remote Procedure Call)~\cite{wu2022bring}, MQTT~\cite{hewa2022fog}, LTE-A (Long-Term Evolution Advanced)~\cite{xu2021deep}, OpenFlow~\cite{friha2020robust}, and other. 

\noindent \textbf{Evaluation} studies the methodologies employed within the paper to assess the efficacy of the proposed solutions and approaches. Evaluation methods play a crucial role in validating research findings and ensuring rigor. Examples of evaluation methods include theoretical frameworks, such as formal verification techniques~\cite{Zhou9915530Service,Xu9508194Transaction}, simulation-based analyses~\cite{Nguyen9916263Latency,JAYAKUMAR20221795}, rigorous measurement~\cite{xu2022mudfl,Taskou9846939Blockchain,SAKTHI202273}, system testing methodologies~\cite{Guo9805797Hybrid,Xu9787357Blockchain}, and other empirically-driven assessments. 

\noindent \textbf{Security} refers to the safeguarding of a system against breaches, data loss, damage, and unauthorized access, ensuring its robustness and trustworthiness. Techniques employed to enhance security encompass authentication~\cite{HALGAMUGE2022109402,wang2022dag}, encryption~\cite{baniata2022prifob,Guan9360653Blockchain}, access control~\cite{wang2022dag,electronics11172710}, verifiability~\cite{RAZAQUE20221,baniata2022prifob}, trust management~\cite{Ruggeri2022Innovative,yuan2022jora}, integrity~\cite{nguyen2021bedgehealth,wang2022dag}, confidentiality~\cite{wang2022dag,nguyen2021bedgehealth}, authorization~\cite{wang2022dag}. 

\noindent \textbf{Privacy} refers to the safeguarding of personal data to prevent their disclosure. It ensures that personal and sensitive information of individuals or organizations is kept confidential and is not shared or used without explicit consent of data owners. Techniques employed in the reviewed literature to ensure privacy encompass anonymity~\cite{MASUD202245}, differential privacy~\cite{WANG2022109206}, homomorphic encryption~\cite{Xu1007Privacy}, ring signature~\cite{mrissa2022privacy}, blind signatures~\cite{Liao9145846Blockchain}, secret sharing~\cite{Nguyen9500648Utility} and autoencoders~\cite{kumar2021dbtp2sf}. 

\noindent \textbf{Sustainability} refers to the Sustainable Development Goals (SDGs) outlined by the United Nations\footnote{The 17 Sustainable Development GOALS. Available at~\url {https://sdgs.un.org/goals} (last accessed December 2024)}. This dimension studies how blockchain and edge computing solutions contribute to these goals, which may include but not limited to good health and well-being~\cite{masud2022user}, quality of education~\cite{baniata2022prifob}, responsible consumption and production~\cite{shahidinejad2022ultra}, and clean water and sanitation~\cite{wang2022data}.
        
\noindent \textbf{Blockchain} indicates that the paper specifically studies or employs a particular blockchain system or distributed ledger, with examples ranging from smart contract platforms, Ethereum~\cite{baranwal2022bara}, to digital currencies, Bitcoin~\cite{jijin2022smart}, and specialized solutions such as IOTA~\cite{wang2022dag} tailored for IoT use cases, among others.

\noindent \textbf{Type} signifies the particular blockchain type of the system proposed, studied or used. Including a focus on public blockchains~\cite{baniata2022prifob} (labeled as value 0), characterized by decentralization and open access; private blockchains~\cite{nartey2022blockchain} (labeled as 1) that restrict access to specific entities; hybrid blockchains~\cite{xu2022mudfl} (labeled as 2) that incorporate features from both public and private systems. Lastly, the exploration of consortium blockchains~\cite{nguyen2021bedgehealth} (labeled as 3), where collaborative efforts among multiple organizations play a central role in network management.

\noindent \textbf{Permission} indicates whether the paper contributes to, studies, or relies on a permissioned (0)~\cite{baniata2022prifob,HALGAMUGE2022109402} or permissionless (1)~\cite{wang2022dag,masud2022user} blockchain system. In a permissioned blockchain system, access to the network and participation in consensus mechanisms are restricted to specific entities, often controlled by a central authority or consortium. On the other hand, a permissionless blockchain system allows anyone to join the network, participate in consensus, and validate transactions without needing authorization. Papers categorized under permissioned blockchain systems typically focus on scenarios where controlled access and centralized governance are preferred, while those categorized under permissionless blockchain systems explore decentralized and open-access environments.

\noindent \textbf{Consensus} identifies the consensus protocol proposed, studied, or used in the paper. Consensus protocols, such as Proof of Work (PoW)~\cite{guo2022blockchain}, Proof of Stake (PoS)~\cite{lei2022gbrm}, Proof of Authority (PoA)~\cite{baniata2022prifob}, Proof of Credit (PoC)~\cite{xu2022mudfl}, and Byzantine Fault Tolerance (BFT)~\cite{kumar2022bdedge}, are crucial in blockchain networks, governing how network participants collectively agree on and validate transactions. The consensus protocol indicates the underlying mechanisms that ensure the integrity and agreement of transactions within the network.

\noindent \textbf{Chain} identifies the type of blockchain storage utilized, studied, or proposed in the studied papers. It distinguishes between data storage mechanisms that are off-chain (0)~\cite{baniata2022prifob,xu2022mudfl}, on-chain (1)~\cite{yuan2022jora,nartey2022blockchain}, or both (2)~\cite{alghamdi2022blockchain,han2022blockchain}. While the blockchain serves as a distributed ledger shared among all network nodes, certain applications may opt for off-chain storage due to limitations in storage capacity and cost considerations inherent in blockchain technology. Off-chain data storage stores information outside the blockchain in a separate database. 

\noindent \textbf{Reward} serves as the incentivization structure in blockchain systems, motivating network participants to honestly follow a protocol. Incentives can take various forms, including monetary (0)~\cite{poongodi2022novel,bhattacharya20226blocks}, reputation-based (1)~\cite{abdel2022privacy,li2022edgewatch}, hybrid (2)~\cite{guo2022incentive,xu2022mudfl}, and multi-dimensional (3)~\cite{yuan2021csedge,olivares2019novel}. Monetary rewards entail nodes receiving tokens or coins with economical values, while reputation-based rewards have an underlying value related to past behavior and activity of the nodes. A hybrid incentive mechanism combines both reputation and monetary rewards for successful task execution. Multi-dimensional incentives involve the use of multiple values/currencies, such as a token combining carbon emissions and reputation.

\section{Methodology}\label{sec:method}

\begin{figure}[!htbp]
\centering
\includegraphics[clip, trim=5.4cm 15.5cm 5.2cm 2cm, width=\columnwidth]{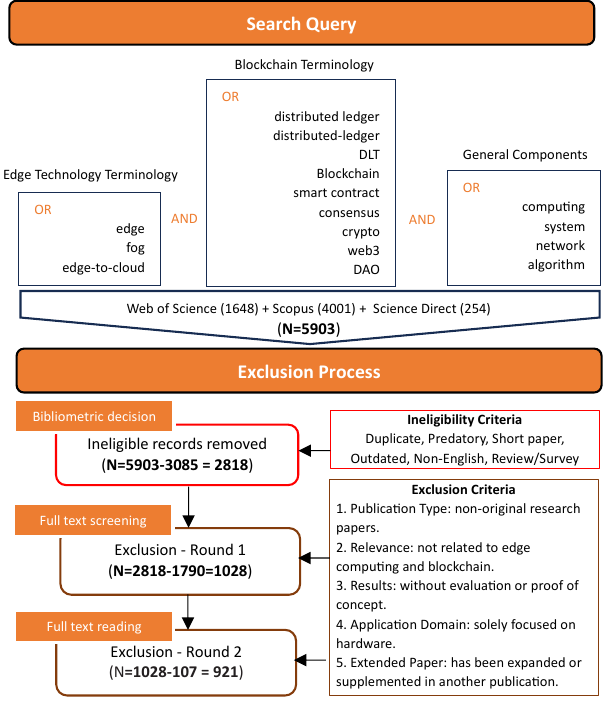}
\caption{Identification and selection process of studies for classification, as detailed in Section~\ref{sec:method}. This comprehensive and rigorous methodology brings together two large fields—edge computing and blockchain. Starting with a total of 5,903 studies from three widely used databases, the subsequent exclusion and review procedures refine the selection to the final set of 921 articles for analysis.}
\label{fig:meth}
\end{figure}

This systematic review adheres to the PRISMA Guidelines~\cite{page2021prisma} for meta-analyses and reporting systematic reviews. The detailed protocol of the review process is available in the GitHub Repository\footnote{\url{https://github.com/TDI-Lab/Edge-Blockchain-Nexus} last accessed June 2025}. Figure~\ref{fig:meth} provides an overview of the search query and study selection procedure conducted across three prominent academic databases: Scopus, ScienceDirect, and Web of Science~\cite{lorenz2023systematic, gusenbauer2020academic}. Following the exclusion steps, the final sample size comprises N = 921 articles for consideration. This substantial number stands in contrast to the existing smaller-scale studies~\cite{alzoubi2023blockchain,alhumam2023cyber,khan2023security,alzoubi2022systematic}, enabling a systematic quantitative meta-analysis of various study dimensions, including causal models to elucidate the interplay between the two studied computing paradigms.

\subsection{Search and Selection Methodology}

Articles included in our study were published within the timeframe spanning from January 2015 to October 2022, ensuring relevance to contemporary advancements in the field. Other articles published in the last 3 years confirm the key trends captured within this chosen time period. Further papers are not included to keep avoid bloating this large-scale study, while keeping it concise and with high standards of rigor. Our search query was meticulously designed to capture a broad spectrum of articles pertinent to our review scope. Notably, we filtered out editorials, book chapters, books, non-English papers, and non-peer-reviewed articles during the search process to maintain a high-quality dataset. The designed query yielded a total of N = 5903 articles across a search scope including titles, abstracts, and full texts. Following systematic screening, encompassing criteria such as predatory journal identification\footnote{BEALL'S LIST. Available at \url{https://beallslist.net/}, last accessed June 2025}, separation of review papers, language restrictions (non-English papers are removed), page length threshold (\textless{} 5-page papers are removed), ensuring papers are published within the determined time range, and duplicate removal, we retained 2818 records for further analysis.

\par Following this, we conducted two rounds of exclusions, overseen by separate reviewers possessing strong knowledge of the area. Each round involved five reviewers (authors of this paper) independently assessing the exclusion criteria to mitigate potential biases. The exclusion criteria encompass: (i) Publication type: Exclusion of articles categorized as reviews or surveys, ensuring inclusion of original research papers. (ii) Relevance: Determination of whether the paper aligns with one or both study perspectives. (iii) Results: Evaluation of whether the study contains an assessment or proof of concept. (iv) Application domain: Examination to ensure the study focuses on software rather than hardware applications. 
During the initial round, a full-text screening was undertaken when pertinent information could not be ascertained solely from the abstracts. Subsequently, in the second round, a thorough review of the full texts was conducted. Following both rounds of screening, a total of 921 articles were selected for inclusion in the final review sample.

\begin{figure}
    \includegraphics[width=\linewidth]
    {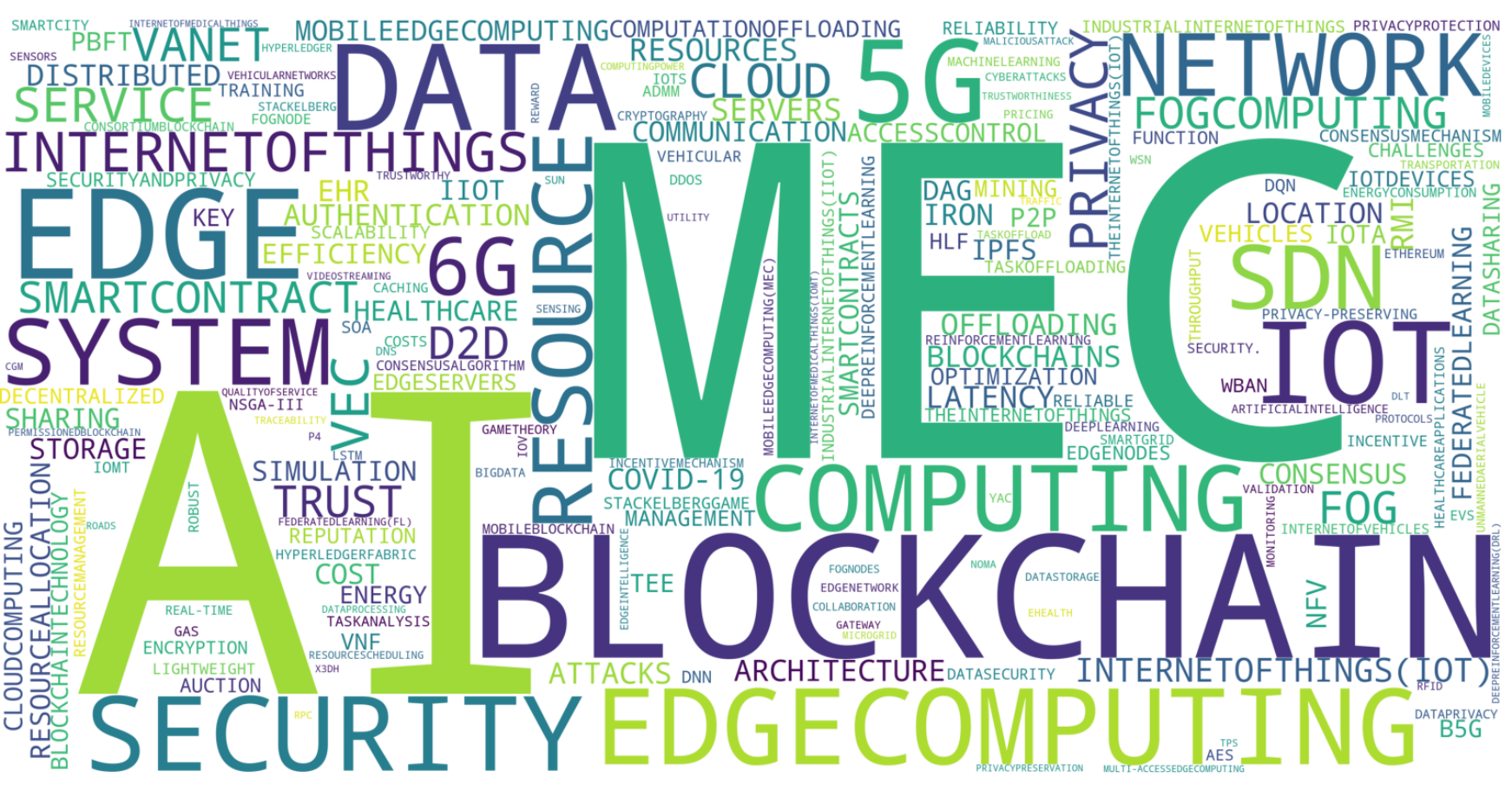}
    \caption{Key themes and focal points within the interface of edge computing and blockchain, derived from the abstracts and keywords of studies reviewed in the SLR.}
    \label{fig:wordcloud}
\end{figure}

Figure~\ref{fig:wordcloud} visualizes the most frequently occurring terms extracted from the abstracts and keywords of the 921 papers reviewed in this SLR. The prominence of terms such as `blockchain', `computing', `edge', `network', `system', `AI', `data', `security', and IoT underscores the primary areas of interest and study. Additionally, smaller yet significant terms including `vehicle', `cost', `cloud', and `distributed', `storage', `offloading', `resource', `management', `smart contract', `consensus', `location', `latency', `architecture', `privacy', `authentication', `trust', highlight critical subtopics and specific issues being addressed. 

\subsection{Meta-analysis}
Following the comprehensive search and selection process, we identified 921 relevant papers, on which we conduct five types of data-driven analysis aiming at answering the research questions. 

\subsubsection{Comparative Analysis: Exploring Blockchain and Edge Computing Interaction Patterns}
The comparative analysis is conducted to explore the differences in applications, problems, design choices, and research maturity across the three perspectives (see Section \ref{sec:perspective}). We categorize the selected papers into these three perspectives. Then, we meticulously compared these categories based on the extracted data, highlighting key differences in application domains, the problems addressed, the design choices made, and the maturity levels of the research.

\subsubsection{Descriptive Analysis: Exploring Dimension Characteristics Over Time}
The descriptive analysis aims to summarize the overall characteristics of the literature, including growth trends, focus areas, and key features. We conducted a statistical analysis of the dataset, summarizing the frequency and distribution of various attributes such as application domains, addressed problems, and technologies used. This analysis provides answer to the research question of how the two computer paradigms co-evolve using the statistical data of the dimensions over time.

\subsubsection{Multiple Correspondence Analysis: Identify Study Patterns}
To identify key research themes in the synergy between blockchain and edge computing, Multiple Correspondence Analysis (MCA) is employed to reduce the dimensionality of the categorical data, revealing the primary dimensions that capture significant variance in the literature. We transformed the categorical data into a binary matrix to identify principal components (dimensions) and their associated eigenvalues. The outcome of the MCA provides a grouping of a number of research patterns from the large amount of papers.

\section{Results}\label{sec:results}
This section answers the research questions using the meta-analysis results of the 921 papers reviewed over the 22 dimensions. 

\subsection{Blockchain and Edge Computing Interaction}\label{sec:qualresult}

To understand how research on blockchain and edge computing interacts to create solutions for different problems and applications, the three perspectives (see Section~\ref{sec:perspective}) are used to categorize the papers for further analysis. 

\paragraph{Blockchain-assisted edge computing systems are more prominently studied than edge-assisted blockchains}
Overall, 75\% of the total reviewed papers focus on blockchain solutions within edge computing environments (Perspective 1), highlighting the pronounced emphasis on leveraging blockchain to address challenges in edge computing. Conversely, the perspective of employing edge computing infrastructure to tackle blockchain challenges (Perspective 2) constitutes 19\% of the total papers; papers that address both edge computing and blockchain (Perspective 3) are found in 6\% of the papers. 

The number of methodological papers increases in   Perspective 1 and Perspective 3 over time. Both perspectives have no methodological papers before 2019, while in 2022, Perspective 1 has 7\% of methodological papers, and Perspective 3 has 12.5\%. However, Perspective 2 shows much fewer methodological papers, with only 2 papers over the five years. For Perspectives 1 and 3, this growing interest in developing methodologies indicates a higher research maturity compared to Perspective 2.

Another dimension, the TRL level also demonstrates that edge-assisted blockchain systems are less technically mature. We distinguish low TRL levels of 1, 2, and 3 and higher levels of 4, 5, 6. In Perspective 1, 30\% of papers have high TRL levels; similarly, in Perspective 3, 29\% of papers have high TRL levels. However, Perspective 2 has a relatively smaller percentage of 25\%. Notably, there is a progression towards high TRL levels for all three perspectives. In total, the percentage increases from 15\% in 2018 to 30\% in 2022, indicating a growing technological maturity in the synergy of blockchain and edge computing. 

\begin{figure*}
    \centering
    \includegraphics[width=\linewidth]{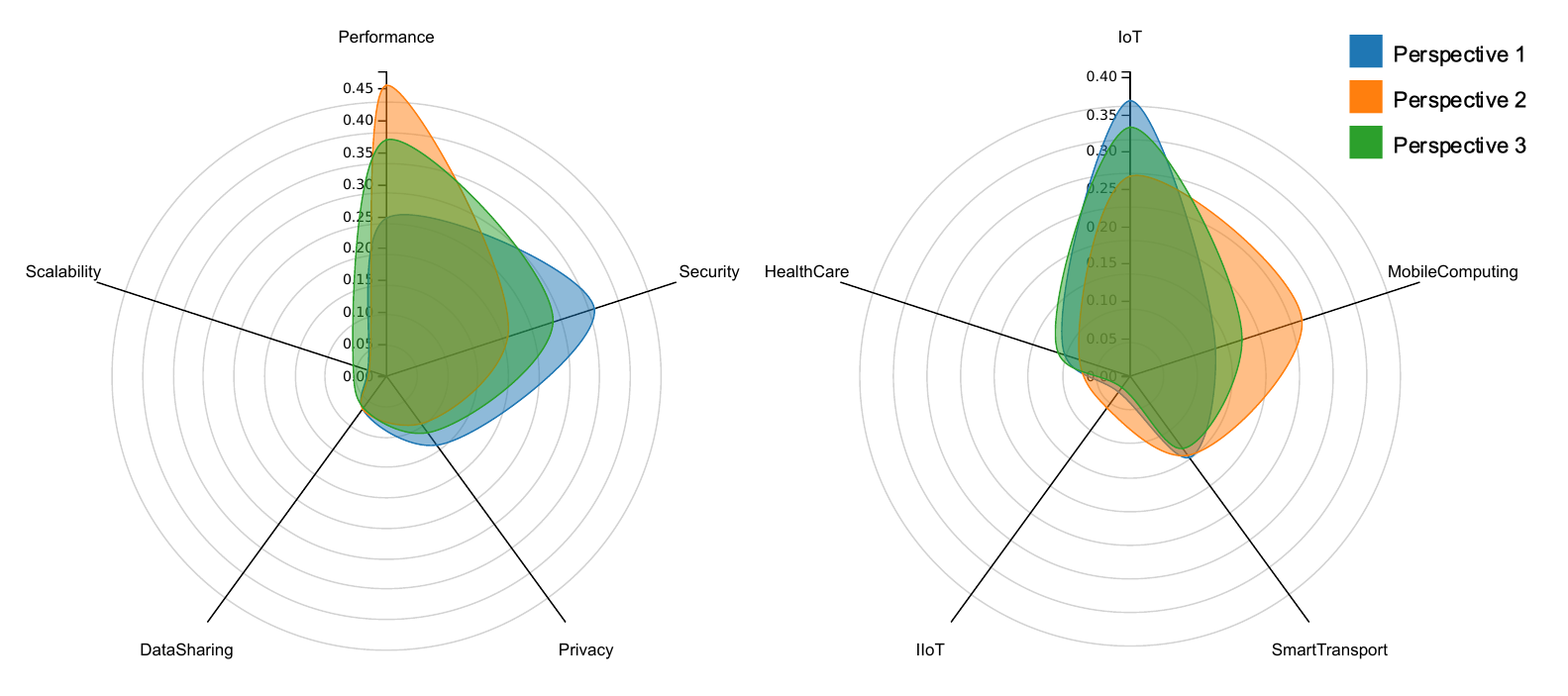}
    \caption{Top five problems (left) and applications (right) studied in the three perspectives, shown in percentage of papers}
    \label{fig:perspetiveRadar}
    \vspace{-10pt}
\end{figure*}

\paragraph{Blockchain for edge computing security; edge computing for blockchain performance}
The studied problems and applications reveal subtle differences in the motivation of combining blockchain and edge computing from different perspectives. As shown in Figure \ref{fig:perspetiveRadar}, the primary problem addressed in Perspective 1 is security, indicating that blockchain technology is predominantly utilized to enhance the security of edge computing systems. Conversely, in Perspectives 2 and 3, edge computing is mainly leveraged to improve the performance of blockchain systems as the studied problems suggest. 

Application-wise, a stronger focus on mobile computing is found exclusively in Perspective 2 compared to Perspective 1 and 3. Papers in this category (Perspective 2 and mobile computing) propose to use mobile edge networks to provide performance advancements to blockchains~\cite{zuo2021computation, liang2022resource, zhang2022optimal}. On the other hand, under the same application category but different perspectives, the introduction of blockchain to assist edge computing in mobile computing provides security advancements~\cite{ma2022optimized, wang2022applying} and an incentive layer to encourage mobile user participation~\cite{chen2022blockchain}. 

\paragraph{Improving blockchain and edge performance via AI methods for resource allocation}
AI-assisted resource allocation increases performance in the nexus of the two computing paradigm~\cite{liu2021virtual}. In Figure~\ref{fig:sankey}b we explore the resource allocation designs and how machine learning is used to achieve resource allocations. It is observed that resource allocation techniques and AI methods combined are predominately solving performance problems, where mobile computing shows a high prevalence under the strong performance constraints of mobile phones. 50\% of this challenge is solved through the strategic approach of offloading. Upon further exploration of offloading solutions, our analysis reveals that reinforcement learning approaches play a pivotal role~\cite{jiang2020intelligent, xu2021blockchain}, featured in 45\% of papers dedicated to implementing computational offloading schemes. Following closely, heuristic~\cite{yuan2022jora, cheng2021trusted} and supervised~\cite{almalki2022enabling, wang2022smart} solutions emerge as the second and third most prominent approaches, each accounting for nearly 20\%.

\subsection{Co-evolution of Blockchain and Edge Computing Nexus}\label{sec:descresult}

The studied problems and solutions of the two computing paradigms are studied over the passage of time to understand the co-evolution of the system design choices and which blockchain systems interact with edge computing. 

\subsubsection{Evolution of the key challenges}
Regarding the problems the papers address, Figure~\ref{fig:probsus}a shows that performance, privacy, and security stand out as the predominant problems covering 76\% of the studied problems. Notably, privacy demonstrates the most significant increase in the last two years, with a 74\% relative increase from 2020 (8.4\%) to 2021 (14.6\%). 

The rising challenges in privacy are potentially related to the key applications studied in the papers (Figure~\ref{fig:probsus}b). For example, mobile computing comes with privacy concerns as it involves mobile phones that contain sensitive data from users~\cite{li2021bpt}; similarly, smart transportation and healthcare require specific awareness of privacy when processing personal data from individuals~\cite{chen2021privacy, wu2021blockchain}. These three applications have drawn increasing attention: together, they account for 38\% of papers in 2018 to 47\% of papers in 2022. Specifically, the most significant growing trend is shown in smart transport applications, covering from only 5\% of the papers in 2018 to 14\% in 2022.

\paragraph{Increasing focus on data confidentiality and privacy}
Protecting data confidentiality against adversaries in such distributed systems where there are multiple sources of sensitive data is the highlighted challenge. Figure~\ref{fig:probsus}c dives into the most studied problem, security, and illustrates the various security challenges explored in the reviewed literature. Authentication emerges as the predominant area of focus that also exhibits a steady upward trend in its significance over time, from 25\% in 2019 to 39\% in 2022. Similarly, access control is observed to follow a comparable trajectory, ranking as the second most investigated security domain in 13\% of the papers dedicated to security studies. 

Bringing anonymity to edge nodes is a popular edge computing privacy approach provided by blockchains. Figure~\ref{fig:probsus}d shows approximately 40\% of the privacy enhancing strategies involved in the papers incorporate anonymity, which is provided in most blockchain systems. When applied to edge computing, edge nodes can be represented as asymmetric key pairs in the network without revealing their identity. In addition to anonymity, encryption is another frequently employed method, used in 25\% of the privacy studies to protect privacy by ensuring data confidentiality. Notably, differential privacy, being a privacy-enhancing technology with low computational cost designed to process data from multiple data sources, demonstrates a significant uptrend, from 6.8\% of the papers in 2020 to 18.5\% in 2022. 

\subsubsection{Evolution of Blockchain Design Choices}

\begin{figure*}
    \subfloat[Problems]{\includegraphics[width=0.33\linewidth]{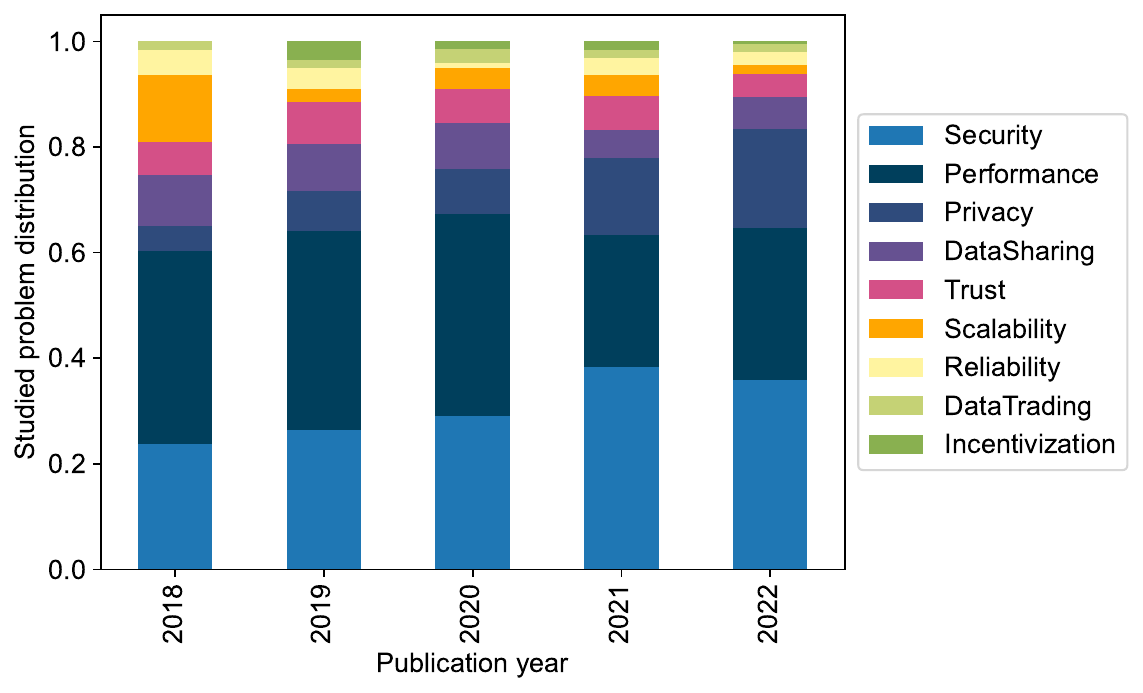}}\hfill
    \subfloat[Applications]{\includegraphics[width=0.33\linewidth]{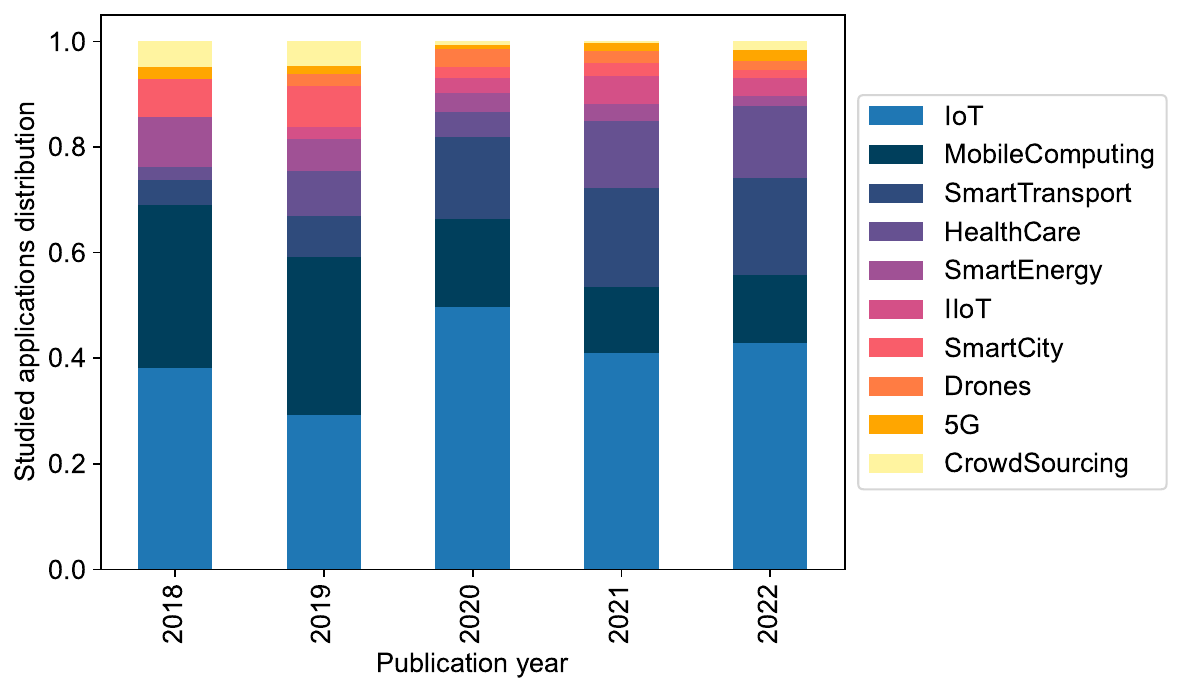}}\hfill
    \subfloat[Security]{\includegraphics[width=0.33\linewidth]{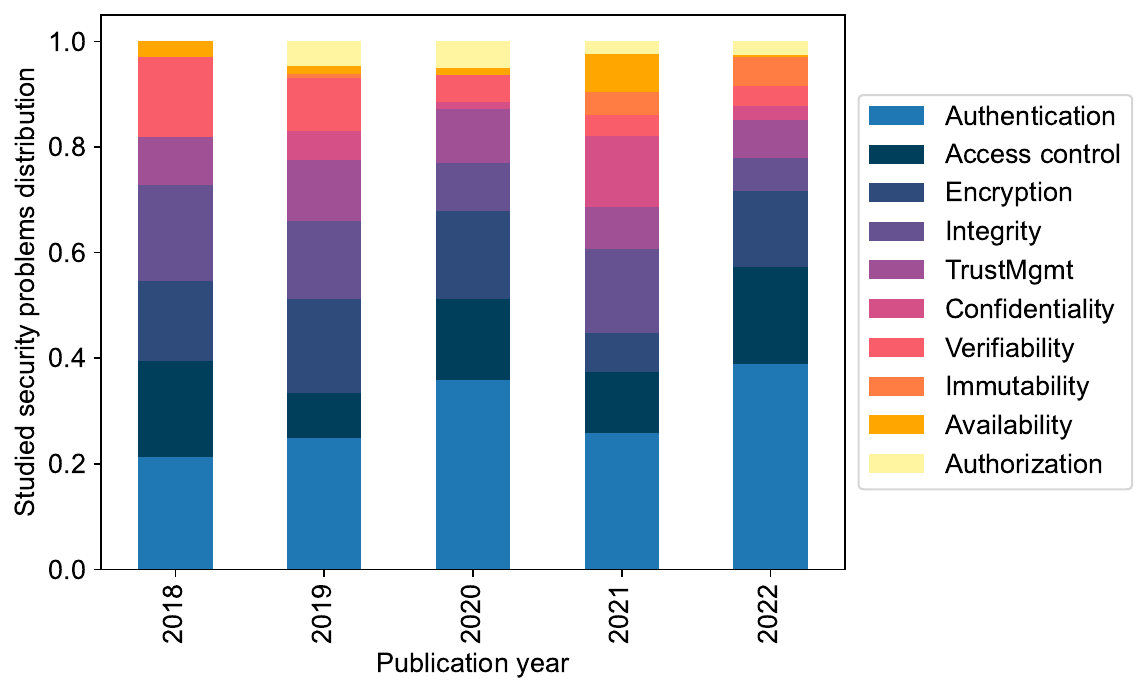}}\hfill
    \\
    \subfloat[Privacy mechanisms]{\includegraphics[width=0.33\linewidth]{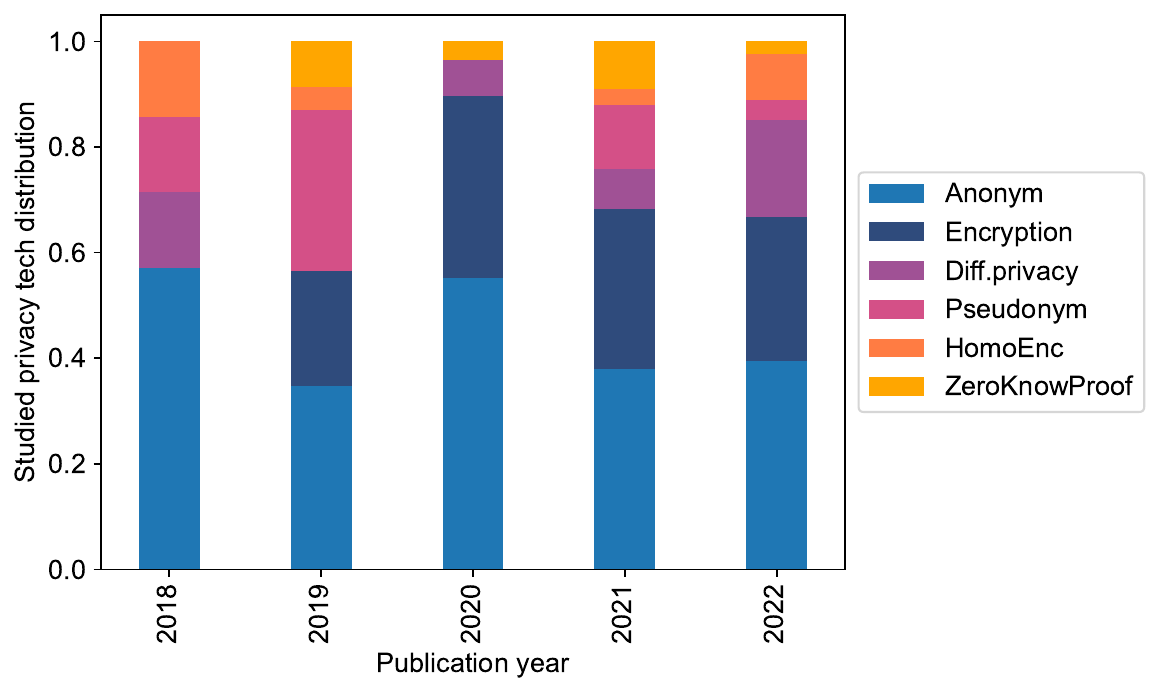}}\hfill
    \subfloat[Consensus]{\includegraphics[width=0.33\linewidth]{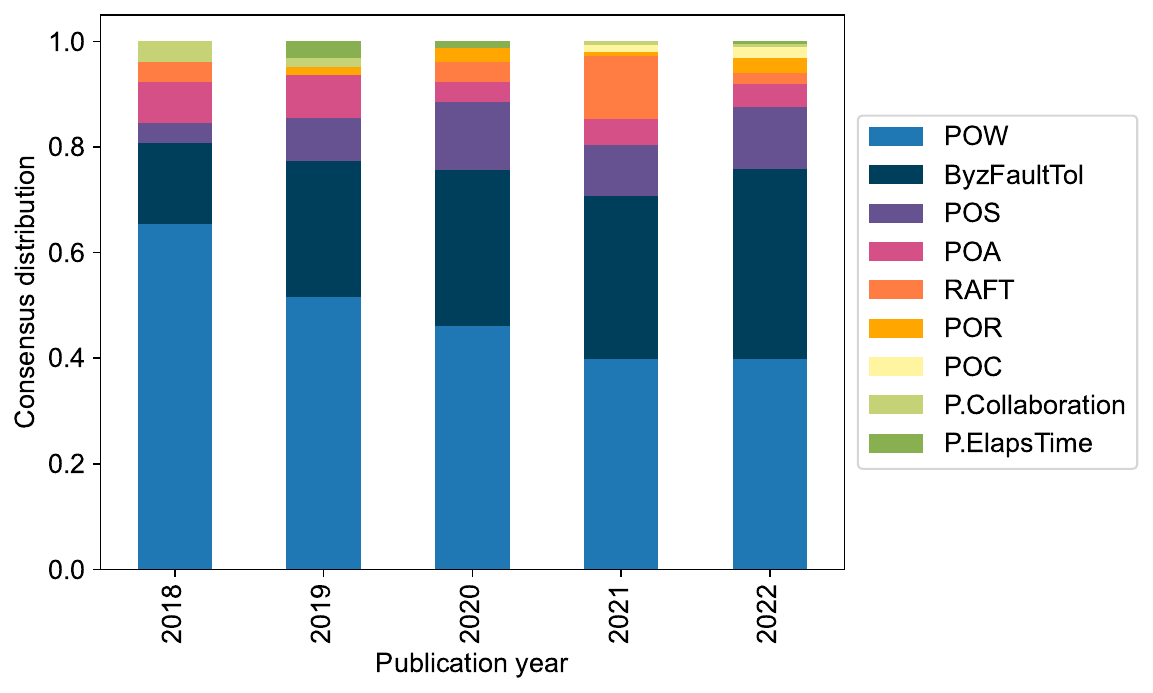}}\hfill
    \subfloat[Reward mechanisms]{\includegraphics[width=0.33\linewidth]{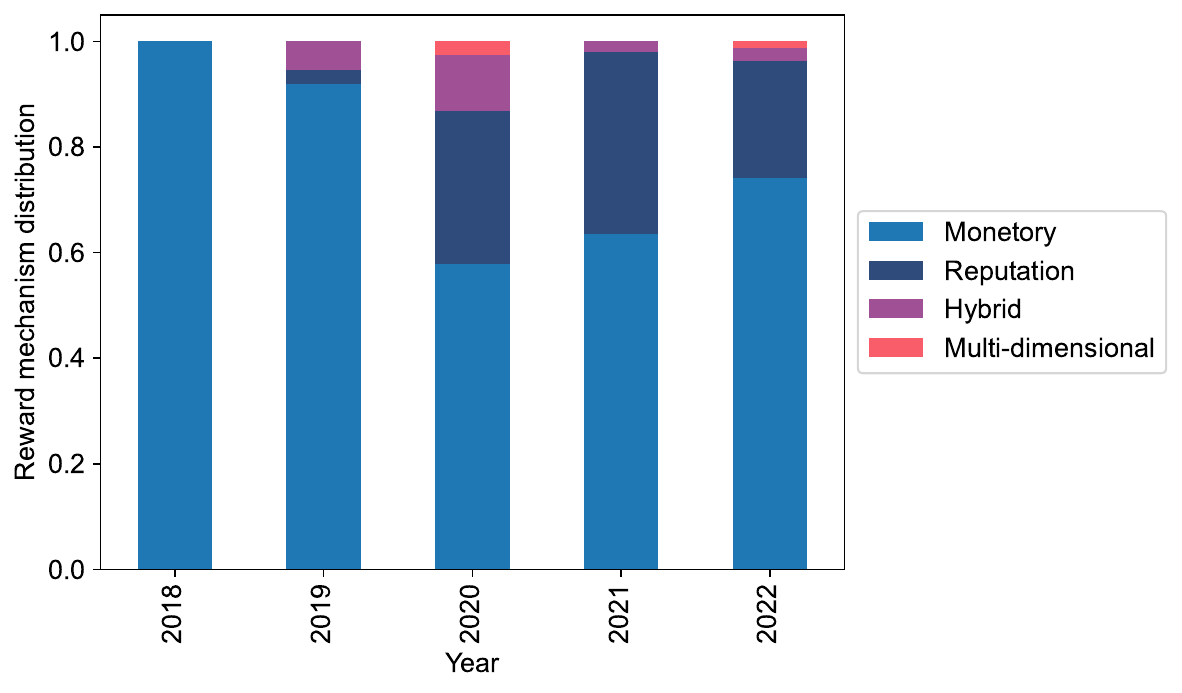}}\hfill
    
    \caption{How the dimension attributes in 95\% of data studied evolve over time}
    \label{fig:probsus}
\end{figure*}

\begin{figure*}
    \subfloat[Studied problems and their association with blockchain design features]{\includegraphics[width=8cm]{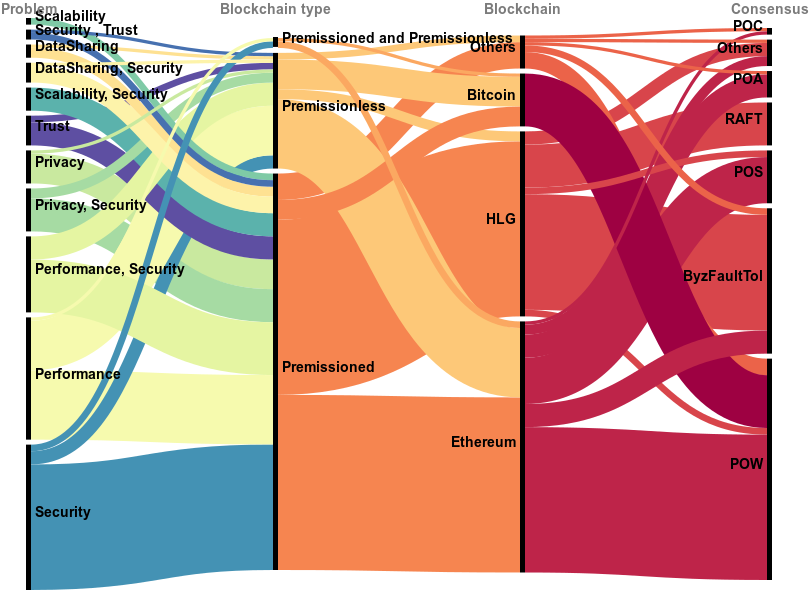}}\hfill
    \subfloat[Applications of studied problems and their association with resource allocation problems and the AI method that tackles them]{\includegraphics[width=8cm]{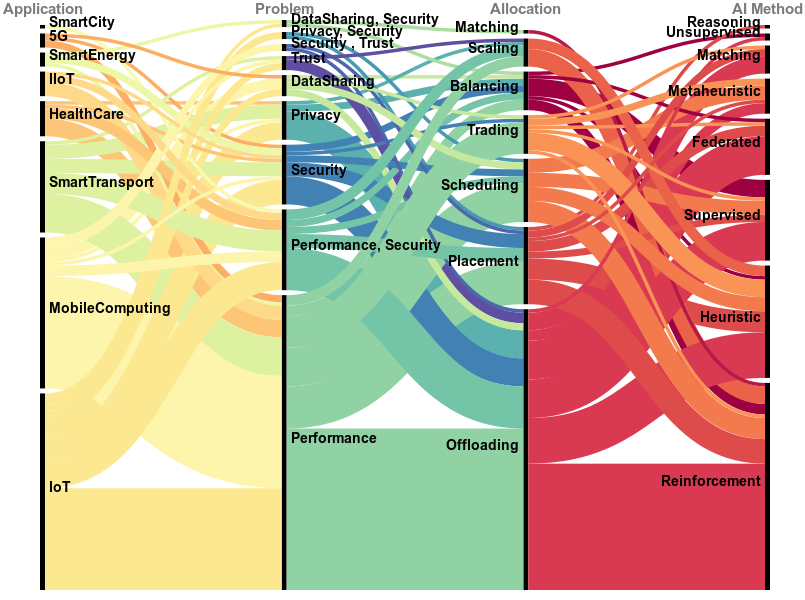}}
    
    \caption{Interplay of design dimensions in the literature on blockchain and edge computing}
    \label{fig:sankey}
\end{figure*}

Regarding design choices of consensus protocols, Figure~\ref{fig:probsus}e shows that the attention on Proof of Work has also gradually diminished over time. It drops to less than 50\% of all consensus algorithms per year since 2019 and further declines to less than 38\% by 2022. This shift potentially relates to the two major disadvantages of Proof of Work, being environmentally unfriendly and inefficient in processing transactions. That said, Proof of Work (PoW), and Proof of Stake (PoS) together still accounted for over 50\% of the total reviewed mechanisms. However, among the papers that specified a blockchain type, only a minority of 22\% runs in a permissionless setting, while 75\% runs in a permissioned setting (see Figure\ref{fig:sankey}a). This contradiction indicates that a systematic guideline for selecting the most suitable consensus mechanism for different scenarios is yet to be developed.

The integration of rewards and incentivization mechanisms in the blockchain-edge computing domain demonstrates a progressive evolution from 4.5\% in 2018 to 36\% in 2022 of all studies reviewed (Figure~\ref{fig:probsus}f). Initially, in 2018, the predominant form of reward was monetary, reflecting a traditional approach to incentivization. However, we observe a notable shift towards more diversified and sophisticated reward systems as the domination of monetary inventive decreases from 100\% in 2018 to 74\% in 2022. In 2022, reputation-based incentives account for 22\% of the total rewards studied or proposed, signaling a growing recognition of the importance of trust and social capital within decentralized networks. Another potential reason for this shift is that permissioned networks that are more prominent in edge computing systems, do not always require monetary rewards to incentivize honest behaviors of the nodes to keep the network safe.

Figure~\ref{fig:sankey}a illustrates the interplay of the following studied dimensions: problems, blockchain types, blockchain platforms, and consensus algorithms. A substantial majority, constituting 75\% of the reviewed literature, utilizes permissioned blockchains. Given that Ethereum is the most popular permissionless blockchain with smart contract support, it is still used to conduct experiments for systems with permissioned design for easier setup and smart contract language support. Similarly, the flow from permissioned blockchain to Bitcoin represents permissioned variances of Bitcoin-like networks. Notably, more than 85\% of papers focus on security concerns and all papers with scalability concerns, with a specified blockchain type, opt for permissioned blockchain. A potential reason is that permissioned systems allow stronger security assumptions on the nodes, under which solutions with better scalability can be proposed.

\subsection{The Key Study Patterns}\label{sec:mcaresult}

\begin{figure*}
    \centering
    \includegraphics[width=\linewidth, height=0.55\textheight]{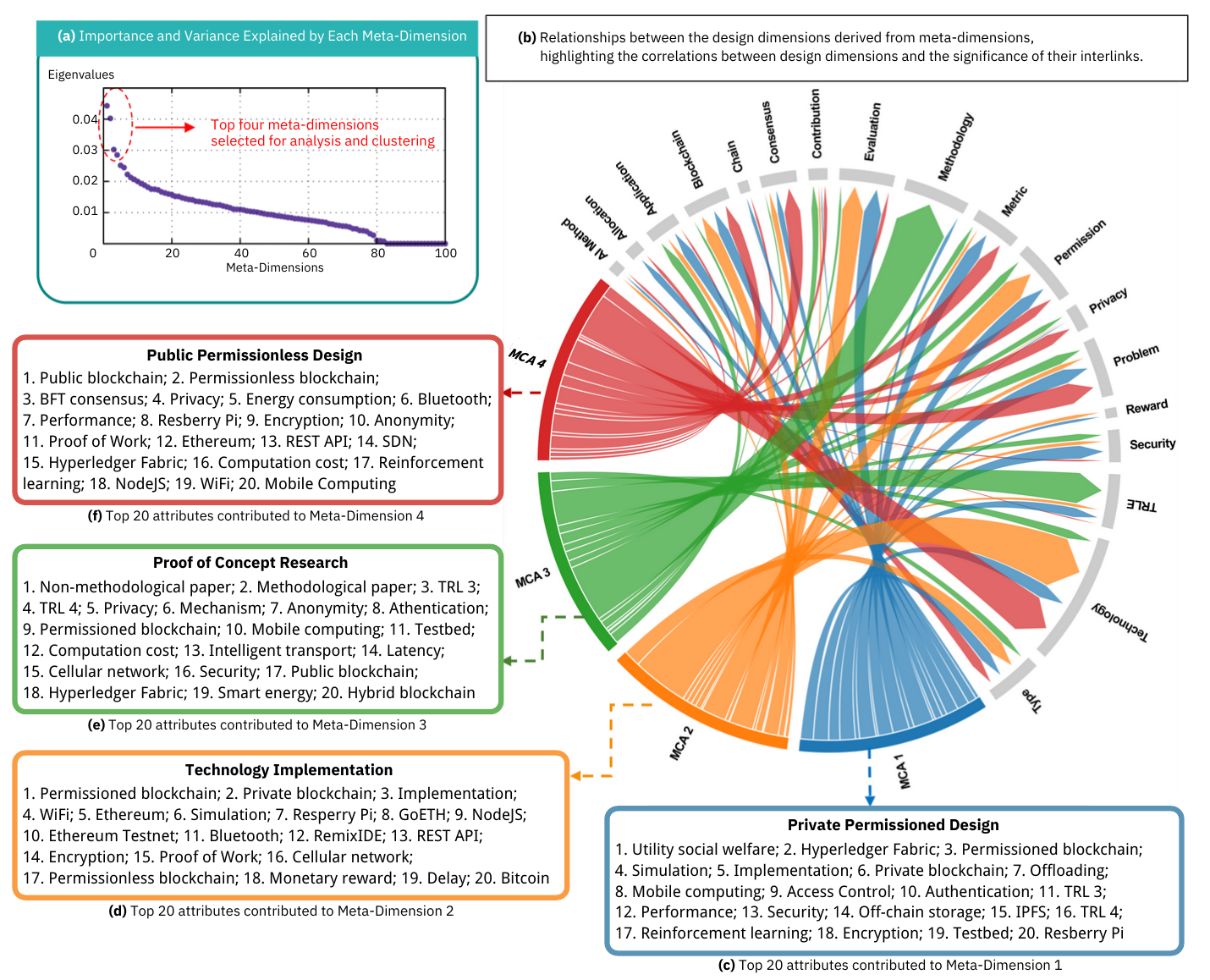}
    \caption{MCA analysis: (a) retained variance across meta-dimensions, (b) design attributes within the top four meta-dimensions, (c-f) top 20 attributes contributed to the four meta-dimensions.}
    \label{fig:infographic}
    \vspace{-10pt}
\end{figure*}

To extract the study patterns and identify the research themes of the reviewed papers, we study the prevalent combinations of dimensions extracted via the MCA analysis. Figure~\ref{fig:infographic}a illustrates the variance explained by different meta-dimensions. Four combinations are sufficient to capture the variance of the data collected from the literature review. Figure~\ref{fig:infographic}b shows how the studied dimensions contribute to defining the top meta-dimensions. Each meta-dimension captures distinct aspects of the literature, revealing relationships among features based on their co-occurrence patterns. 

\noindent \textbf{Pattern 1: Private permissioned design} The first meta-dimension reflects on the design of permissioned blockchain interfacing with edge computing to make secure and efficient systems. It captures the utilization of offloading techniques for performance, and various storage designs such as off-chain storage and IPFS distributed storage protocol to manage data. Security problems including access control and authentication are considered in these studies. The proposed solutions are supported by evaluations through implementation, simulation and testbed analysis.

\noindent \textbf{Pattern 2: Technology implementation}
This meta-dimension primarily captures implementation of blockchain-edge systems using technologies and tools such as Resperry Pi, Bluetooth, NodeJS, and REST API. Both permissioned aand permissionless blockchains involved in the design. Ethereum plays a major role in the implementation and evaluation of the designed systems, supported by the utilization of tools including GoETH client, Ethereum testnet, and RemixIDE, potentially due to the programmability provided by Ethereum Virtual Machine and the Solidity smart contract language. Besides Ethereum, Proof of Work consensus and the Bitcoin network are also found in this pattern.

\noindent \textbf{Pattern 3: Proof of concept research} This meta-dimension encapsulates methodological and non-methodological approaches pivotal to the integration of edge computing and blockchain. TRL 3 and 4 denote a level of maturity in technology development and deployment within this study pattern. It highlights the studies of security and privacy problems in applications such as mobile computing, intelligent transport and smart energy. Various blockchain types are studied in this meta-dimension of basic research for blockchain-edge systems, including permissioned blockchain, hybrid blockchain and public blockchain.

\noindent \textbf{Pattern 4: Public permissionless design} This meta-dimension encompasses a spectrum of attributes pertaining to blockchain technology, consensus mechanisms, privacy considerations, and performance metrics. Its focus lies in optimizing performance while addressing privacy concerns, underscoring the significance of blockchain platforms including Hyperledger and Ethereum, consensus protocols including BFT and Proof of Work, and privacy measures including encryption and anonymity. Furthermore, it incorporates technologies including bluetooth, Resberry Pi, REST API, SDN, NodeJS, and WiFi. Energy consumption and computation cost are used as measurements in this meta-dimension.

\section{Discussion}\label{sec:disc}

The prevalent synergy of blockchain-assisted edge computing for security also demonstrates research of higher TRL levels, while the edge-assisted blockchain systems for performance follow in prevalence and technological readiness, with mobile computing a key application. The challenge of preserving on-chain solutions and security standards in edge computing environments that are usually supported by IoT devices may explain this discrepancy. 

Privacy is a key challenge that motivates the exploration of solutions based on the two computing paradigms. The cryptographic solutions of distributed ledgers along with the more localized processing of data in edge computing encourage the combination of privacy protection techniques with privacy-by-design approaches. On the other hand, more lightweight permissioned blockchain environments remain prevalent in the context of edge computing, while reputation-based incentive mechanisms gain momentum over monetary rewards. 

Using machine learning methods to extract key features that distinguish the conducted research in the interface of blockchain and edge computing reveals four key patterns: one pattern captures the technology-driven research across the two computing paradigms, while another pattern reflects on diverse proof-of-concept research conducted at TRL levels of 3-4 involving permissioned, public and hybrid blockchains. The other two patterns are distinguished by public permissionless designs vs. private and permissioned. As a result, three driving factors determine the development of the two computing paradigms: supported technology, proof of concepts and the openness of system design.

\section{Conclusion and Future Work}\label{sec:conc}

To conclude, the novel large-scale systematic literature review in the nexus of blockchain and edge computing shows that the two paradigms of decentralized computing provide complementary solutions and opportunities to address long-standing challenges such as scalability/efficiency and security/privacy in applications such as mobile computing, smart mobility, Internet of Things applications and Smart Cities. Applying blockchain solutions to edge computing is the prevalent approach over using edge computing for blockchain, with heterogeneous IoT devices posing several security and technical obstacles to support blockchain. System design plays a key, in particular the openness of the distributed ledgers as public permissionless vs. private permissioned. Proof of concepts, methodology and technology are also important criteria that determine the research conducted in the nexus of the computing paradigms.

Future work includes a more thorough analysis of epistemological aspects related to key advancements and technology adoption by collecting additional supportive data such as bibilometrics (paper citations), patents, market analysis and a mapping of industry.  

\section*{Acknowledgements}

This work is funded by a UKRI Future Leaders Fellowship (MR\-/W009560\-/1): \emph{Digitally Assisted Collective Governance of Smart City Commons--ARTIO}'. 

\bibliographystyle{unsrt}
\bibliography{reference}

\appendix

\section*{Additional Table}\label{app:relatedwork}

Table~\ref{tab:lit-rev1} provides a comparison of search queries and databases used in existing literature with those utilized in our study. 

\begin{table*}[htbp]
    \centering
    \caption{Systematic literature review summary}
    \label{tab:lit-rev1}
    \tiny
    \begin{tabular}{p{1.3cm}p{8cm}p{4cm}}
        \toprule
        Study&Search Query&Database\\
        \midrule
        
       (2023)~\cite{alzoubi2023blockchain}&``Edge computing" OR ``Fog computing" AND ``Blockchain" &IEEE Xplore, Elsevier ScienceDirect, Wiley Online Library, SpringerLink, Google Scholar, MDPI Online, SAGE Publication, ACM Digital Library, Emerald Insight\\

       (2023)~\cite{alhumam2023cyber}&(cybersecurity Based Blockchain OR security Based Blockchain OR Blockchain Based Security OR Block-chain based security)AND( FC OR fog computing OR Fog Processing OR fog computing architecture AND IoT) AND (decentralized security threats) AND (cyber-attacks)&Google Scholar, Saudi Digital library\\

       (2023)~\cite{khan2023security}&``Cloud computing security" AND ``QoS in cloud environment", ``Cloud computing security" AND/OR ``trust-based secure cloud models", ``Cloud computing security" AND ``blockchain technology", ``Cloud computing security" AND/OR ``scheduling in cloud computing", ``Cloud computing security" AND ``edge computing in cloud environment"&Scopus, IEEE Xplore, MDPI, Science Direct, Springer, Wiley, Google Scholars\\ 

       (2022)~\cite{alzoubi2022systematic}&``Edge computing" OR ``Fog computing" AND ``Blockchain"&IEEE Xplore, ACM Digital Library, Elsevier ScienceDirect, SpringerLink, Google Scholar, Emerald Insight, Wiley Online Library, SAGE Publication, MDPI Online\\

       This work&(``distributed ledger" OR ``distributed-ledger" OR ``DLT" OR ``blockchain" OR ``smart contract" OR ``consensus" OR ``crypto" OR ``web3" OR ``DAO") AND (``edge" OR ``fog" OR ``edge-to-cloud") AND (``computing" OR ``system" OR ``network" OR ``algorithm")& Scopus, ScienceDirect, Web of Science\\
 \bottomrule
    \end{tabular}
\end{table*}

\clearpage

\printglossary

\end{document}